\documentclass[12pt]{article}

\usepackage{graphics,graphpap, myfeynarts}

\newcommand{\bea}{\begin{eqnarray}}
\newcommand{\eea}{\end{eqnarray}}
\newcommand{\beq}{\begin{equation}}
\newcommand{\eeq}{\end{equation}}

\def\msbar{\ifmmode{\overline{\rm MS}} \else{$\overline{\rm MS}$} \fi}
\def\drbar{\ifmmode{\overline{\rm DR}} \else{$\overline{\rm DR}$} \fi}
\def\sf{\ifmmode{\tilde{f}} \else{$\tilde{f}$} \fi}
\def\st{\ifmmode{\tilde{t}} \else{$\tilde{t}$} \fi}
\def\sb{\ifmmode{\tilde{b}} \else{$\tilde{b}$} \fi}
\def\sq{\ifmmode{\tilde{q}} \else{$\tilde{q}$} \fi}
\def\sg{\ifmmode{\tilde{g}} \else{$\tilde{g}$} \fi}
\def\bbar{\ifmmode{\bar{b}} \else{$\bar{b}$} \fi}
\def\tbar{\ifmmode{\bar{t}} \else{$\bar{t}$} \fi}
\def\qbar{\ifmmode{\bar{q}} \else{$\bar{q}$} \fi}
\def\ksla{{k \hspace{-2mm} \slash}}

\def\sq              {{\tilde q}}

\def\sf              {{\tilde f}}

\newcommand\bvec{\left( \begin{array}{c}}
\newcommand\evec{\end{array}\right)}

\newcommand\bmat{\left( \begin{array}{cc}}
\newcommand\emat{\end{array}\right)}

\newcommand\ch{{\tilde{\chi}}}
\renewcommand\d{\delta}

\def\ksla{{k \hspace{-2.2mm} \slash}}

\newcommand\cchpm{{\tilde\chi^\pm}}
\newcommand\cchmp{{\tilde\chi^\mp}}

\newcommand\cch{{\tilde\chi^0}}

\def\chp             {\tilde \chi^+}

\def\onehf           {{\textstyle \frac{1}{2}}}

\renewcommand\d{\delta}
\newcommand\D{\Delta}
\renewcommand\a{\alpha}
\renewcommand\b{\beta}
\newcommand\g{\gamma}
\newcommand\e{\epsilon}

\newcommand{\ZN}{N}

\def\su{\ifmmode{\tilde{u}} \else{$\tilde{u}$} \fi}
\def\sd{\ifmmode{\tilde{d}} \else{$\tilde{d}$} \fi}

\def\non             {\nonumber}

\begin{document}

\pagestyle{empty} \vspace*{-1cm}
\begin{flushright}
  HEPHY-PUB 785/04 \\
  hep-ph/0402134
\end{flushright}

\vspace*{1.4cm}

\begin{center}
\begin{Large} \bf
Full one-loop corrections to neutralino pair production in {\boldmath $e^+e^-$} annihilation
\end{Large}

\vspace{10mm}

{\large W.~\"Oller, H.~Eberl, W.~Majerotto}

\vspace{6mm}
\begin{tabular}{l}
 {\it Institut f\"ur Hochenergiephysik der \"Osterreichischen
 Akademie der Wissenschaften,}\\
{\it A--1050 Vienna, Austria}
\end{tabular}

\vspace{20mm}

\begin{abstract}
We present the full one-loop radiative corrections to pair production of neutralinos in $e^+e^-$ collisions within the Minimal 
Supersymmetric Standard Model. Particular attention is paid to the definition of weak and QED corrections. 
The non-universal QED corrections are extracted by subtracting the initial state radiation. 
We give numerical results for two different SUSY scenarios for $e^+e^-\rightarrow\cch_1\cch_2$ and
$e^+e^-\rightarrow\cch_2\cch_2$. The weak and QED corrections are up to several
percent or even higher and need to be taken into account at future linear collider experiments.  
\end{abstract}
\end{center}

\vfill

\newpage
\pagestyle{plain} \setcounter{page}{2}

\section{Introduction}
\vspace{2mm} In the Minimal Supersymmetric Standard Model (MSSM) \cite{MSSM}, one has two charginos $\cchpm_1$ and $\cchpm_2$, which
are the fermion mass eigenstates of the supersymmetric partners of the $W^\pm$ and the charged Higgs states $H^\pm_{1,2}$.
Likewise, there are four neutralinos $\cch_1$-$\cch_4$, which are the fermion mass eigenstates of the supersymmetric partners
of the photon, the $Z^0$ boson, and the neutral Higgs bosons $H^0_{1,2}$. Their mass matrix
depends on the parameters $M$, $M'$, $\mu$, and $\tan\b$, where $M$ and $M'$ are the SU(2) and U(1) gauge mass parameter, and
$\tan\b=\frac{v_2}{v_1}$ with $v_{1,2}$ the vacuum expectation values of the two neutral Higgs doublet fields. 
If supersymmetry is realized in nature, 
charginos and neutralinos should be found in the next generation of high energy experiments at Tevatron, LHC and a future $e^+e^-$
collider, . Especially at a linear $e^+e^-$ collider, it will be possible to perform
measurements with high precision \cite{tesla, lincol}. In particular, it has been shown in \cite{tesla} that the masses of
charginos and neutralinos can be measured within an accuracy of $\Delta m_{\tilde{\chi}^{\pm,0}} = 0.1 - 1$ GeV. It is therefore
obvious that such a high precision requires equally accurate theoretical predictions. Despite the complexity, for some SUSY
processes the full one-loop corrections have already been calculated: for $e^+e^-\rightarrow\cchpm_i\cchmp_j$, $i,j=1,2$, in \cite{BlankHollik},
for $e^+e^-\rightarrow\tilde{l}_i\,{\bar{\!\tilde{l}}}_j$, $l=e, \mu$, $i,j=L,R$ in \cite{Freitas}, 
$e^+e^-\rightarrow\tilde{f}_i\ \, {\bar{\!\!\tilde{f}}}_{\!\!j}$, 
$f=q, l, \nu$ (including the third generation) in \cite{ArhribHollik, KarolChris}. As to decays, the full one-loop corrections were 
calculated for $\tilde{q}_i\rightarrow q\cch_i$, $i=1-4$, and $\tilde{q}_i\rightarrow q\cchpm_k$, $k=1,2$, in \cite{ghs}, and for the
decays $A^0\rightarrow \tilde{f}_1\ {\bar{\!\!\tilde{f}}}_{\!2}$, $\tilde{f}_2\rightarrow \tilde{f}_1A_0$ in \cite{ChrisA0}, where $A^0$
is the pseudoscalar Higgs particle. All these calculations have shown that the corrections are important for precise predictions of
cross sections, branching ratios and asymmetries.\\
In this paper, we present the calculation of the complete one-loop corrections to the neutralino production
$e^+e^-\rightarrow\cch_i\cch_j$, $i,j=1-4$.
\\ \qquad For the calculation of higher order corrections, renormalization of the MSSM is necessary. For this purpose, one has to
employ appropriate renormalization conditions, or equi\-valently, one has to fix the counter terms for the SUSY parameters. In this
paper, we adopt the on-shell scheme for the chargino and neutralino system of \cite{mcvienna}. Equivalent methods were developed 
in \cite{mchollik, ghs}. The schemes only differ in the fixing of the counter terms of the parameters $M$, $M'$ and $\mu$. Hence the meaning of these
parameters is different at loop-level. The schemes, however, yield the same results for observables as masses, cross sections,
widths, etc. . \\
Starting from the tree-level in section \ref{treelevel}, we outline the calculation of the one-loop corrections in
section \ref{corrections} discussing the renormalization both of the SUSY and SM parameters. The process-independent corrections to
the neutralino mass matrix are included in an improved tree-level. Particular attention is paid to a proper definition of the weak and QED 
corrections as they latter play an important r\^{o}le. In section \ref{numericalresults}, we represent a detailed numerical analysis for
$e^+e^-\rightarrow\cch_1\cch_2$ and $e^+e^-\rightarrow\cch_2\cch_2$ for a higgsino and a gaugino scenario for $\cch_1$ and $\cch_2$.
Conclusions are given in section \ref{conclusions}.

\section{Tree-level}\label{treelevel}
In the MSSM the neutralino sector is specified by the gaugino mass parameters $M$ and $M'$, 
the higgsino mass parameter $\mu$ and the Higgs mixing angle $\tan\beta$, all appearing in the neutralino mass matrix
(in the bino, $W^3$-ino, $H_{1,2}$-ino basis)
\begin{equation}
    Y = \left(\begin{array}{cccc}
M' & 0 & - m_Z \sin\theta_W \cos\beta & m_Z \sin\theta_W
\sin\beta\\ 0 & M & m_Z \cos\theta_W \cos\beta & -m_Z \cos\theta_W
\sin\beta\\ - m_Z \sin\theta_W \cos\beta & m_Z \cos\theta_W
\cos\beta & 0 & -\mu\\ m_Z \sin\theta_W \sin\beta & -m_Z
\cos\theta_W \sin\beta & -\mu & 0
\end{array}\right) \,.
 \label{neumat1}
\end{equation}
With the unitary matrix $\ZN$, which diagonalizes the mass matrix $Y$
\begin{equation}
\mbox{diag}(m_{\tilde\chi_1^0},\,m_{\tilde\chi_2^0},\,m_{\tilde\chi_3^0},\,m_{\tilde\chi_4^0})= \ZN^* Y \ZN^\dag 
\, , 
\end{equation}
we can rotate from the gauge eigenstates $\tilde{\psi}_j^0={(-i\tilde{\lambda}',-i\tilde{\lambda}^3,\tilde{\psi}_{H_1}^1,\tilde{\psi}_{H_2}^2)}_j$ to
the neutralino mass eigenstate basis $\cch_i=\ZN_{ij}\tilde{\psi}^0_j$. 

\begin{figure}[h!]
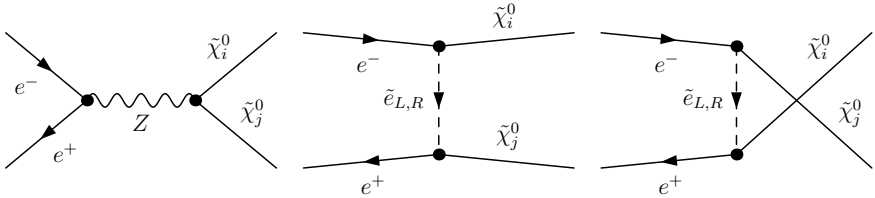

\begin{center}
\mbox{\resizebox{200mm}{!}{
\begin{feynartspicture}(750,150)(4,1)
\FADiagram{}
\FAProp(0.,15.)(6.,10.)(0.,){/Straight}{1}
\FALabel(2.48771,11.7893)[tr]{$e^-$}
\FAProp(0.,5.)(6.,10.)(0.,){/Straight}{-1}
\FALabel(3.51229,6.78926)[tl]{$e^+$}
\FAProp(20.,15.)(14.,10.)(0.,){/Straight}{0}
\FALabel(16.6478,13.0187)[br]{$\tilde \chi_i^0$}
\FAProp(20.,5.)(14.,10.)(0.,){/Straight}{0}
\FALabel(17.3522,8.01869)[bl]{$\tilde \chi_j^0$}
\FAProp(6.,10.)(14.,10.)(0.,){/Sine}{0}
\FALabel(10.,8.93)[t]{$Z$}
\FAVert(6.,10.){0}
\FAVert(14.,10.){0}

\FADiagram{}
\FAProp(0.,15.)(10.,14.)(0.,){/Straight}{1}
\FALabel(4.84577,13.4377)[t]{$e^-$}
\FAProp(0.,5.)(10.,6.)(0.,){/Straight}{-1}
\FALabel(5.15423,4.43769)[t]{$e^+$}
\FAProp(20.,15.)(10.,14.)(0.,){/Straight}{0}
\FALabel(14.8706,15.3135)[b]{$\tilde \chi_i^0$}
\FAProp(20.,5.)(10.,6.)(0.,){/Straight}{0}
\FALabel(15.1294,6.31355)[b]{$\tilde \chi_j^0$}
\FAProp(10.,14.)(10.,6.)(0.,){/ScalarDash}{1}
\FALabel(8.93,10.)[r]{$\tilde e_{L,R}$}
\FAVert(10.,14.){0}
\FAVert(10.,6.){0}

\FADiagram{}
\FAProp(0.,15.)(10.,14.)(0.,){/Straight}{1}
\FALabel(4.84577,13.4377)[t]{$e^-$}
\FAProp(0.,5.)(10.,6.)(0.,){/Straight}{-1}
\FALabel(5.15423,4.43769)[t]{$e^+$}
\FAProp(20.,15.)(10.,6.)(0.,){/Straight}{0}
\FALabel(16.98,13.02)[br]{$\tilde \chi_i^0$}
\FAProp(20.,5.)(10.,14.)(0.,){/Straight}{0}
\FALabel(17.52,8.02)[bl]{$\tilde \chi_j^0$}
\FAProp(10.,14.)(10.,6.)(0.,){/ScalarDash}{1}
\FALabel(9.03,10.)[r]{$\tilde e_{L,R}$}
\FAVert(10.,14.){0}
\FAVert(10.,6.){0}

\end{feynartspicture}
}}
\caption[fig1]
{Tree--Level}
 \label{fig:tree}
\end{center}
\end{figure}

At tree-level and neglecting the electron mass in all Yukawa couplings the production process
\bea
e^+e^- \rightarrow \cch_i\cch_j  \qquad \qquad \qquad (i,j = 1,2,3,4)  \nonumber
\eea
contains contributions from the Feynman diagrams shown in Fig.~\ref{fig:tree}: The direct $s$--channel due to the $Z^0$ exchange
and the crossed $t$-- and $u$--channel due to the $\tilde{e}_{L, R}$ exchanges. 

From the interaction Lagrangian
\begin{eqnarray}
{\cal L}_{Z^0 \bar{e} e}&=&-\frac{g}{\cos\theta_W}Z_\mu^0\bar{e}\gamma^\mu[C_LP_L+C_RP_R]e,\\
{\cal L}_{Z^0 \cch_i\cch_j}&=&\frac{g}{2\cos\theta_W}Z_\mu^0\bar{\cch_i}\gamma^\mu[O_{ij}^{''L}P_L+O_{ij}^{''R}P_R]\cch_j,\\
{\cal L}_{e\tilde{e}\cch_i}&=&gf_i^L\bar{e}P_R\cch_i\tilde{e}_L+gf_i^R\bar{e}P_L\cch_i\tilde{e}_R+\rm{h.c.},
\end{eqnarray} 
we obtain the couplings
\begin{eqnarray}
C_{L,R} &=& I^{3L,R} + \sin^2\theta_W, \quad I^{3L}=-\frac{1}{2}, \quad I^{3R}=0,\\ 
O_{ij}^{''L}&=&-O_{ij}^{''R*}=-\frac{1}{2}\ZN_{i3}\ZN_{j3}^*+\frac{1}{2}\ZN_{i4}\ZN_{j4}^*,\\
f_i^L&=&-\frac{\sqrt{2}}{2}\left( \tan\theta_W\ZN_{i1}+\ZN_{i2}\right),\quad f_i^R=\sqrt{2}\tan\theta_W\ZN^*_{i1}.
\end{eqnarray}

\section{One-loop corrections}\label{corrections}
\begin{figure}[h!]
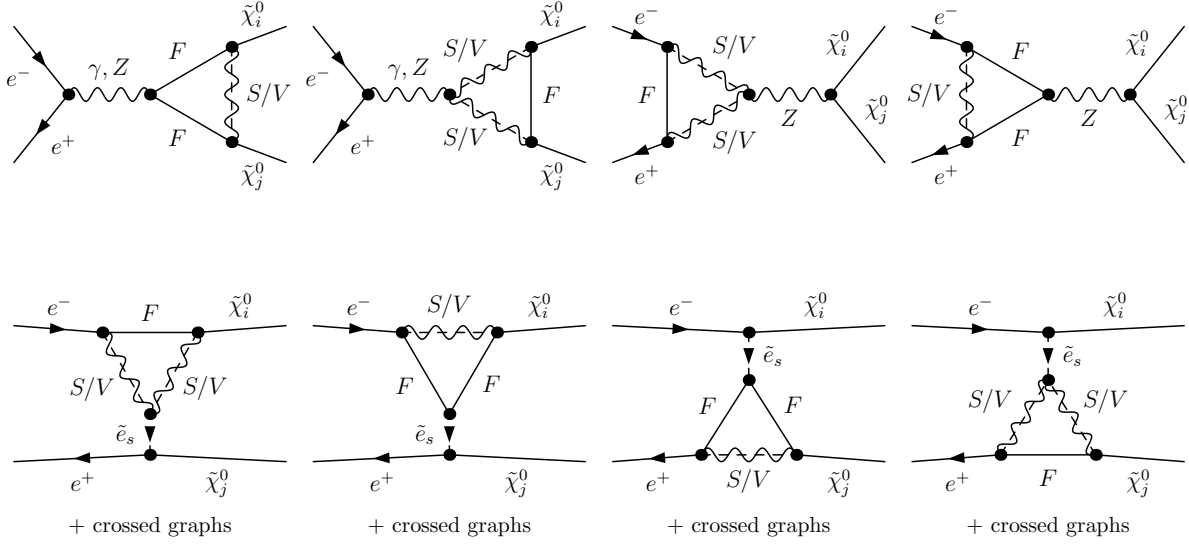

\hspace{-2cm}
\mbox{\resizebox{200mm}{!}{
\begin{feynartspicture}(750,300)(4,2)
\FADiagram{}
\FAProp(0.,15.)(4.,10.)(0.,){/Straight}{1}
\FALabel(1.26965,12.0117)[tr]{$e^-$}
\FAProp(0.,5.)(4.,10.)(0.,){/Straight}{-1}
\FALabel(2.73035,7.01172)[tl]{$e^+$}
\FAProp(20.,15.)(16.,13.5)(0.,){/Straight}{0}
\FALabel(17.5435,14.9872)[b]{$\tilde \chi_i^0$}
\FAProp(20.,5.)(16.,6.5)(0.,){/Straight}{0}
\FALabel(17.5435,5.01277)[t]{$\tilde \chi_j^0$}
\FAProp(4.,10.)(10.,10.)(0.,){/Sine}{0}
\FALabel(7.,11.07)[b]{$\gamma,Z$}
\FAProp(16.,13.5)(16.,6.5)(0.,){/ScalarDash}{0}
\FAProp(16.,13.5)(16.,6.5)(0.,){/Sine}{0}
\FALabel(17.07,10.)[l]{$S/V$}
\FAProp(16.,13.5)(10.,10.)(0.,){/Straight}{0}
\FALabel(12.699,12.6089)[br]{$F$}
\FAProp(16.,6.5)(10.,10.)(0.,){/Straight}{0}
\FALabel(12.699,7.39114)[tr]{$F$}
\FAVert(4.,10.){0}
\FAVert(16.,13.5){0}
\FAVert(16.,6.5){0}
\FAVert(10.,10.){0}

\FADiagram{}
\FAProp(0.,15.)(4.,10.)(0.,){/Straight}{1}
\FALabel(1.26965,12.0117)[tr]{$e^-$}
\FAProp(0.,5.)(4.,10.)(0.,){/Straight}{-1}
\FALabel(2.73035,7.01172)[tl]{$e^+$}
\FAProp(20.,15.)(16.,13.5)(0.,){/Straight}{0}
\FALabel(17.5435,14.9872)[b]{$\tilde \chi_i^0$}
\FAProp(20.,5.)(16.,6.5)(0.,){/Straight}{0}
\FALabel(17.5435,5.01277)[t]{$\tilde \chi_j^0$}
\FAProp(4.,10.)(10.,10.)(0.,){/Sine}{0}
\FALabel(7.,11.07)[b]{$\gamma,Z$}
\FAProp(16.,13.5)(16.,6.5)(0.,){/Straight}{0}
\FALabel(16.82,10.)[l]{$F$}
\FAProp(16.,13.5)(10.,10.)(0.,){/ScalarDash}{0}
\FAProp(16.,13.5)(10.,10.)(0.,){/Sine}{0}
\FALabel(12.825,12.3929)[br]{$S/V$}
\FAProp(16.,6.5)(10.,10.)(0.,){/ScalarDash}{0}
\FAProp(16.,6.5)(10.,10.)(0.,){/Sine}{0}
\FALabel(12.825,7.60709)[tr]{$S/V$}
\FAVert(4.,10.){0}
\FAVert(16.,13.5){0}
\FAVert(16.,6.5){0}
\FAVert(10.,10.){0}

\FADiagram{}
\FAProp(0.,15.)(4.,13.5)(0.,){/Straight}{1}
\FALabel(2.54424,15.2213)[b]{$e^-$}
\FAProp(0.,5.)(4.,6.5)(0.,){/Straight}{-1}
\FALabel(2.54424,4.77869)[t]{$e^+$}
\FAProp(20.,15.)(16.,10.)(0.,){/Straight}{0}
\FALabel(17.4649,12.8321)[br]{$\tilde \chi_i^0$}
\FAProp(20.,5.)(16.,10.)(0.,){/Straight}{0}
\FALabel(18.5351,7.8321)[bl]{$\tilde \chi_j^0$}
\FAProp(16.,10.)(10.,10.)(0.,){/Sine}{0}
\FALabel(13.,8.93)[t]{$Z$}
\FAProp(4.,13.5)(4.,6.5)(0.,){/Straight}{0}
\FALabel(3.18,10.)[r]{$F$}
\FAProp(4.,13.5)(10.,10.)(0.,){/ScalarDash}{0}
\FAProp(4.,13.5)(10.,10.)(0.,){/Sine}{0}
\FALabel(7.301,12.6089)[bl]{$S/V$}
\FAProp(4.,6.5)(10.,10.)(0.,){/ScalarDash}{0}
\FAProp(4.,6.5)(10.,10.)(0.,){/Sine}{0}
\FALabel(7.301,7.39114)[tl]{$S/V$}
\FAVert(4.,13.5){0}
\FAVert(4.,6.5){0}
\FAVert(16.,10.){0}
\FAVert(10.,10.){0}

\FADiagram{}
\FAProp(0.,15.)(4.,13.5)(0.,){/Straight}{1}
\FALabel(2.54424,15.2213)[b]{$e^-$}
\FAProp(0.,5.)(4.,6.5)(0.,){/Straight}{-1}
\FALabel(2.54424,4.77869)[t]{$e^+$}
\FAProp(20.,15.)(16.,10.)(0.,){/Straight}{0}
\FALabel(17.4649,12.8321)[br]{$\tilde \chi_i^0$}
\FAProp(20.,5.)(16.,10.)(0.,){/Straight}{0}
\FALabel(18.5351,7.8321)[bl]{$\tilde \chi_j^0$}
\FAProp(16.,10.)(10.,10.)(0.,){/Sine}{0}
\FALabel(13.,8.93)[t]{$Z$}
\FAProp(4.,13.5)(4.,6.5)(0.,){/Sine}{0}
\FAProp(4.,13.5)(4.,6.5)(0.,){/ScalarDash}{0}
\FALabel(2.93,10.)[r]{$S/V$}
\FAProp(4.,13.5)(10.,10.)(0.,){/Straight}{0}
\FALabel(7.301,12.6089)[bl]{$F$}
\FAProp(4.,6.5)(10.,10.)(0.,){/Straight}{0}
\FALabel(7.301,7.39114)[tl]{$F$}
\FAVert(4.,13.5){0}
\FAVert(4.,6.5){0}
\FAVert(16.,10.){0}
\FAVert(10.,10.){0}

\FADiagram{+ crossed graphs}
\FAProp(0.,15.)(6.5,14.5)(0.,){/Straight}{1}
\FALabel(3.36888,15.8154)[b]{$e^-$}
\FAProp(0.,5.)(10.,5.5)(0.,){/Straight}{-1}
\FALabel(5.0774,4.18193)[t]{$e^+$}
\FAProp(20.,15.)(13.5,14.5)(0.,){/Straight}{0}
\FALabel(16.6503,15.5662)[b]{$\tilde \chi_i^0$}
\FAProp(20.,5.)(10.,5.5)(0.,){/Straight}{0}
\FALabel(14.9351,4.43162)[t]{$\tilde \chi_j^0$}
\FAProp(10.,5.5)(10.,8.5)(0.,){/ScalarDash}{-1}
\FALabel(8.93,7.)[r]{$\tilde e_s$}
\FAProp(6.5,14.5)(13.5,14.5)(0.,){/Straight}{0}
\FALabel(10.,15.32)[b]{$F$}
\FAProp(6.5,14.5)(10.,8.5)(0.,){/ScalarDash}{0}
\FAProp(6.5,14.5)(10.,8.5)(0.,){/Sine}{0}
\FALabel(7.39114,11.199)[tr]{$S/V$}
\FAProp(13.5,14.5)(10.,8.5)(0.,){/ScalarDash}{0}
\FAProp(13.5,14.5)(10.,8.5)(0.,){/Sine}{0}
\FALabel(12.3929,11.325)[tl]{$S/V$}
\FAVert(6.5,14.5){0}
\FAVert(10.,5.5){0}
\FAVert(13.5,14.5){0}
\FAVert(10.,8.5){0}

\FADiagram{+ crossed graphs}
\FAProp(0.,15.)(6.5,14.5)(0.,){/Straight}{1}
\FALabel(3.36888,15.8154)[b]{$e^-$}
\FAProp(0.,5.)(10.,5.5)(0.,){/Straight}{-1}
\FALabel(5.0774,4.18193)[t]{$e^+$}
\FAProp(20.,15.)(13.5,14.5)(0.,){/Straight}{0}
\FALabel(16.6503,15.5662)[b]{$\tilde \chi_i^0$}
\FAProp(20.,5.)(10.,5.5)(0.,){/Straight}{0}
\FALabel(14.9351,4.43162)[t]{$\tilde \chi_j^0$}
\FAProp(10.,5.5)(10.,8.5)(0.,){/ScalarDash}{-1}
\FALabel(8.93,7.)[r]{$\tilde e_s$}
\FAProp(6.5,14.5)(13.5,14.5)(0.,){/Sine}{0}
\FAProp(6.5,14.5)(13.5,14.5)(0.,){/ScalarDash}{0}
\FALabel(10.,15.57)[b]{$S/V$}
\FAProp(6.5,14.5)(10.,8.5)(0.,){/Straight}{0}
\FALabel(7.39114,11.199)[tr]{$F$}
\FAProp(13.5,14.5)(10.,8.5)(0.,){/Straight}{0}
\FALabel(12.3929,11.325)[tl]{$F$}
\FAVert(6.5,14.5){0}
\FAVert(10.,5.5){0}
\FAVert(13.5,14.5){0}
\FAVert(10.,8.5){0}

\FADiagram{+ crossed graphs}
\FAProp(0.,15.)(10.,14.5)(0.,){/Straight}{1}
\FALabel(5.0774,15.8181)[b]{$e^-$}
\FAProp(0.,5.)(6.5,5.5)(0.,){/Straight}{-1}
\FALabel(3.36888,4.18457)[t]{$e^+$}
\FAProp(20.,15.)(10.,14.5)(0.,){/Straight}{0}
\FALabel(14.9351,15.5684)[b]{$\tilde \chi_i^0$}
\FAProp(20.,5.)(13.5,5.5)(0.,){/Straight}{0}
\FALabel(16.6503,4.43383)[t]{$\tilde \chi_j^0$}
\FAProp(10.,14.5)(10.,11.)(0.,){/ScalarDash}{1}
\FALabel(11.07,12.75)[l]{$\tilde e_s$}
\FAProp(6.5,5.5)(13.5,5.5)(0.,){/ScalarDash}{0}
\FAProp(6.5,5.5)(13.5,5.5)(0.,){/Sine}{0}
\FALabel(10.,4.43)[t]{$S/V$}
\FAProp(6.5,5.5)(10.,11.)(0.,){/Straight}{0}
\FALabel(7.63324,8.46794)[br]{$F$}
\FAProp(13.5,5.5)(10.,11.)(0.,){/Straight}{0}
\FALabel(12.5777,8.60216)[bl]{$F$}
\FAVert(10.,14.5){0}
\FAVert(6.5,5.5){0}
\FAVert(13.5,5.5){0}
\FAVert(10.,11.){0}

\FADiagram{+ crossed graphs}
\FAProp(0.,15.)(10.,14.5)(0.,){/Straight}{1}
\FALabel(5.0774,15.8181)[b]{$e^-$}
\FAProp(0.,5.)(6.5,5.5)(0.,){/Straight}{-1}
\FALabel(3.36888,4.18457)[t]{$e^+$}
\FAProp(20.,15.)(10.,14.5)(0.,){/Straight}{0}
\FALabel(14.9351,15.5684)[b]{$\tilde \chi_i^0$}
\FAProp(20.,5.)(13.5,5.5)(0.,){/Straight}{0}
\FALabel(16.6503,4.43383)[t]{$\tilde \chi_j^0$}
\FAProp(10.,14.5)(10.,11.)(0.,){/ScalarDash}{1}
\FALabel(11.07,12.75)[l]{$\tilde e_s$}
\FAProp(6.5,5.5)(13.5,5.5)(0.,){/Straight}{0}
\FALabel(10.,4.43)[t]{$F$}
\FAProp(6.5,5.5)(10.,11.)(0.,){/ScalarDash}{0}
\FAProp(6.5,5.5)(10.,11.)(0.,){/Sine}{0}
\FALabel(7.42232,8.60216)[br]{$S/V$}
\FAProp(13.5,5.5)(10.,11.)(0.,){/ScalarDash}{0}
\FAProp(13.5,5.5)(10.,11.)(0.,){/Sine}{0}
\FALabel(12.5777,8.60216)[bl]{$S/V$}
\FAVert(10.,14.5){0}
\FAVert(6.5,5.5){0}
\FAVert(13.5,5.5){0}
\FAVert(10.,11.){0}
\end{feynartspicture}}}
\caption[vertex]
{Generic Vertex Corrections}
 \label{fig:vertex}
\end{figure}

\begin{figure}[h!]
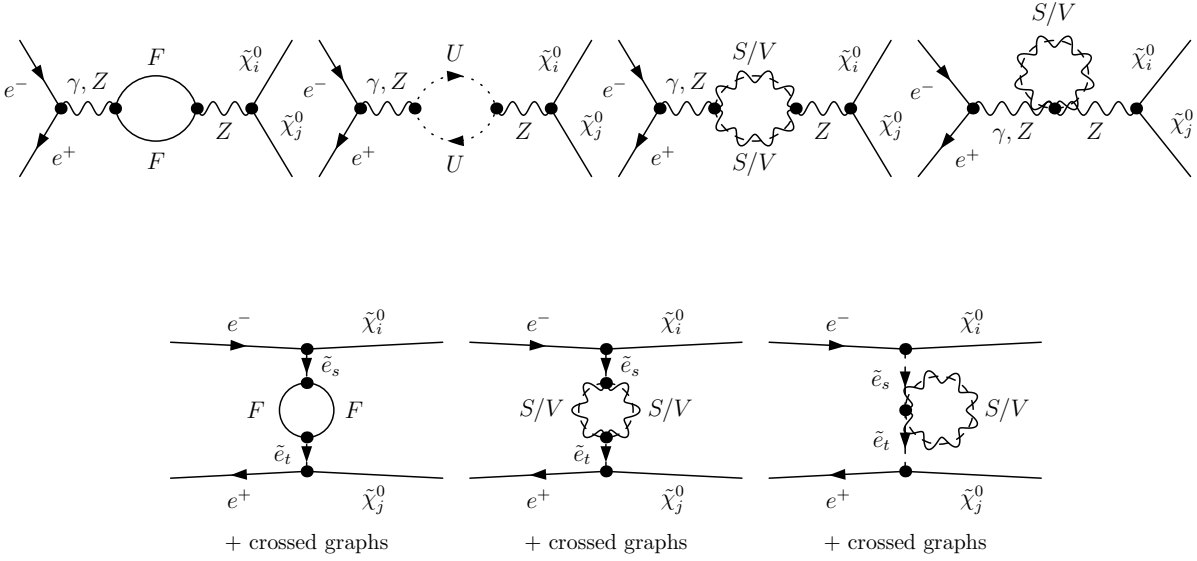

\mbox{\hspace{-2cm}\resizebox{200mm}{!}{

\begin{feynartspicture}(750,150)(4,1)

\FADiagram{}
\FAProp(0.,15.)(3.,10.)(0.,){/Straight}{1}
\FALabel(0.650886,12.1825)[tr]{$e^-$}
\FAProp(0.,5.)(3.,10.)(0.,){/Straight}{-1}
\FALabel(2.34911,7.18253)[tl]{$e^+$}
\FAProp(20.,15.)(17.,10.)(0.,){/Straight}{0}
\FALabel(17.8653,12.6888)[br]{$\tilde \chi_i^0$}
\FAProp(20.,5.)(17.,10.)(0.,){/Straight}{0}
\FALabel(19.1347,7.68884)[bl]{$\tilde \chi_j^0$}
\FAProp(3.,10.)(7.,10.)(0.,){/Sine}{0}
\FALabel(5.,11.07)[b]{$\gamma,Z$}
\FAProp(17.,10.)(13.,10.)(0.,){/Sine}{0}
\FALabel(15.,8.93)[t]{$Z$}
\FAProp(7.,10.)(13.,10.)(0.8,){/Straight}{0}
\FALabel(10.,6.78)[t]{$F$}
\FAProp(7.,10.)(13.,10.)(-0.8,){/Straight}{0}
\FALabel(10.,13.22)[b]{$F$}
\FAVert(3.,10.){0}
\FAVert(17.,10.){0}
\FAVert(7.,10.){0}
\FAVert(13.,10.){0}

\FADiagram{}
\FAProp(0.,15.)(3.,10.)(0.,){/Straight}{1}
\FALabel(0.650886,12.1825)[tr]{$e^-$}
\FAProp(0.,5.)(3.,10.)(0.,){/Straight}{-1}
\FALabel(2.34911,7.18253)[tl]{$e^+$}
\FAProp(20.,15.)(17.,10.)(0.,){/Straight}{0}
\FALabel(17.8653,12.6888)[br]{$\tilde \chi_i^0$}
\FAProp(20.,5.)(17.,10.)(0.,){/Straight}{0}
\FALabel(19.1347,7.68884)[bl]{$\tilde \chi_j^0$}
\FAProp(3.,10.)(7.,10.)(0.,){/Sine}{0}
\FALabel(5.,11.07)[b]{$\gamma,Z$}
\FAProp(17.,10.)(13.,10.)(0.,){/Sine}{0}
\FALabel(15.,8.93)[t]{$Z$}
\FAProp(7.,10.)(13.,10.)(0.8,){/GhostDash}{-1}
\FALabel(10.,6.53)[t]{$U$}
\FAProp(7.,10.)(13.,10.)(-0.8,){/GhostDash}{1}
\FALabel(10.,13.47)[b]{$U$}
\FAVert(3.,10.){0}
\FAVert(17.,10.){0}
\FAVert(7.,10.){0}
\FAVert(13.,10.){0}

\FADiagram{}
\FAProp(0.,15.)(3.,10.)(0.,){/Straight}{1}
\FALabel(0.650886,12.1825)[tr]{$e^-$}
\FAProp(0.,5.)(3.,10.)(0.,){/Straight}{-1}
\FALabel(2.34911,7.18253)[tl]{$e^+$}
\FAProp(20.,15.)(17.,10.)(0.,){/Straight}{0}
\FALabel(17.8653,12.6888)[br]{$\tilde \chi_i^0$}
\FAProp(20.,5.)(17.,10.)(0.,){/Straight}{0}
\FALabel(19.1347,7.68884)[bl]{$\tilde \chi_j^0$}
\FAProp(3.,10.)(7.,10.)(0.,){/Sine}{0}
\FALabel(5.,11.07)[b]{$\gamma,Z$}
\FAProp(17.,10.)(13.,10.)(0.,){/Sine}{0}
\FALabel(15.,8.93)[t]{$Z$}
\FAProp(7.,10.)(13.,10.)(0.8,){/ScalarDash}{0}
\FAProp(7.,10.)(13.,10.)(0.8,){/Sine}{0}
\FALabel(10.,6.78)[t]{$S/V$}
\FAProp(7.,10.)(13.,10.)(-0.8,){/ScalarDash}{0}
\FAProp(7.,10.)(13.,10.)(-0.8,){/Sine}{0}
\FALabel(10.,13.22)[b]{$S/V$}
\FAVert(3.,10.){0}
\FAVert(17.,10.){0}
\FAVert(7.,10.){0}
\FAVert(13.,10.){0}

\FADiagram{}
\FAProp(0.,15.)(4.,10.)(0.,){/Straight}{1}
\FALabel(1.26965,12.0117)[tr]{$e^-$}
\FAProp(0.,5.)(4.,10.)(0.,){/Straight}{-1}
\FALabel(2.73035,7.01172)[tl]{$e^+$}
\FAProp(20.,15.)(16.,10.)(0.,){/Straight}{0}
\FALabel(17.4649,12.8321)[br]{$\tilde \chi_i^0$}
\FAProp(20.,5.)(16.,10.)(0.,){/Straight}{0}
\FALabel(18.5351,7.8321)[bl]{$\tilde \chi_j^0$}
\FAProp(4.,10.)(10.,10.)(0.,){/Sine}{0}
\FALabel(7.,8.93)[t]{$\gamma,Z$}
\FAProp(16.,10.)(10.,10.)(0.,){/Sine}{0}
\FALabel(13.,8.93)[t]{$Z$}
\FAProp(10.,10.)(10.,10.)(10.,15.){/ScalarDash}{0}
\FAProp(10.,10.)(10.,10.)(10.,15.){/Sine}{0}
\FALabel(10.,16.07)[b]{$S/V$}
\FAVert(4.,10.){0}
\FAVert(16.,10.){0}
\FAVert(10.,10.){0}
\end{feynartspicture}}}

\mbox{\resizebox{200mm}{!}{
\begin{feynartspicture}(750,150)(4,1)
\FADiagram{+ crossed graphs}
\FAProp(0.,15.)(10.,14.5)(0.,){/Straight}{1}
\FALabel(5.0774,15.8181)[b]{$e^-$}
\FAProp(0.,5.)(10.,5.5)(0.,){/Straight}{-1}
\FALabel(5.0774,4.18193)[t]{$e^+$}
\FAProp(20.,15.)(10.,14.5)(0.,){/Straight}{0}
\FALabel(14.9351,15.5684)[b]{$\tilde \chi_i^0$}
\FAProp(20.,5.)(10.,5.5)(0.,){/Straight}{0}
\FALabel(14.9351,4.43162)[t]{$\tilde \chi_j^0$}
\FAProp(10.,14.5)(10.,12.)(0.,){/ScalarDash}{1}
\FALabel(11.07,13.25)[l]{$\tilde e_s$}
\FAProp(10.,5.5)(10.,8.)(0.,){/ScalarDash}{-1}
\FALabel(8.93,6.75)[r]{$\tilde e_t$}
\FAProp(10.,12.)(10.,8.)(1.,){/Straight}{0}
\FALabel(6.93,10.)[r]{$F$}
\FAProp(10.,12.)(10.,8.)(-1.,){/Straight}{0}
\FALabel(12.82,10.)[l]{$F$}
\FAVert(10.,14.5){0}
\FAVert(10.,5.5){0}
\FAVert(10.,12.){0}
\FAVert(10.,8.){0}

\FADiagram{+ crossed graphs}
\FAProp(0.,15.)(10.,14.5)(0.,){/Straight}{1}
\FALabel(5.0774,15.8181)[b]{$e^-$}
\FAProp(0.,5.)(10.,5.5)(0.,){/Straight}{-1}
\FALabel(5.0774,4.18193)[t]{$e^+$}
\FAProp(20.,15.)(10.,14.5)(0.,){/Straight}{0}
\FALabel(14.9351,15.5684)[b]{$\tilde \chi_i^0$}
\FAProp(20.,5.)(10.,5.5)(0.,){/Straight}{0}
\FALabel(14.9351,4.43162)[t]{$\tilde \chi_j^0$}
\FAProp(10.,14.5)(10.,12.)(0.,){/ScalarDash}{1}
\FALabel(11.07,13.25)[l]{$\tilde e_s$}
\FAProp(10.,5.5)(10.,8.)(0.,){/ScalarDash}{-1}
\FALabel(8.93,6.75)[r]{$\tilde e_t$}
\FAProp(10.,12.)(10.,8.)(1.,){/ScalarDash}{0}
\FAProp(10.,12.)(10.,8.)(1.,){/Sine}{0}
\FALabel(6.93,10.)[r]{$S/V$}
\FAProp(10.,12.)(10.,8.)(-1.,){/ScalarDash}{0}
\FAProp(10.,12.)(10.,8.)(-1.,){/Sine}{0}
\FALabel(13.07,10.)[l]{$S/V$}
\FAVert(10.,14.5){0}
\FAVert(10.,5.5){0}
\FAVert(10.,12.){0}
\FAVert(10.,8.){0}

\FADiagram{+ crossed graphs}
\FAProp(0.,15.)(10.,14.5)(0.,){/Straight}{1}
\FALabel(5.0774,15.8181)[b]{$e^-$}
\FAProp(0.,5.)(10.,5.5)(0.,){/Straight}{-1}
\FALabel(5.0774,4.18193)[t]{$e^+$}
\FAProp(20.,15.)(10.,14.5)(0.,){/Straight}{0}
\FALabel(14.9351,15.5684)[b]{$\tilde \chi_i^0$}
\FAProp(20.,5.)(10.,5.5)(0.,){/Straight}{0}
\FALabel(14.9351,4.43162)[t]{$\tilde \chi_j^0$}
\FAProp(10.,14.5)(10.,10.)(0.,){/ScalarDash}{1}
\FALabel(8.93,12.25)[r]{$\tilde e_s$}
\FAProp(10.,5.5)(10.,10.)(0.,){/ScalarDash}{-1}
\FALabel(8.93,7.75)[r]{$\tilde e_t$}
\FAProp(10.,10.)(10.,10.)(15.,10.){/ScalarDash}{0}
\FAProp(10.,10.)(10.,10.)(15.,10.){/Sine}{0}
\FALabel(15.82,10.)[l]{$S/V$}
\FAVert(10.,14.5){0}
\FAVert(10.,5.5){0}
\FAVert(10.,10.){0}
\end{feynartspicture}}}

\caption[self]
{Generic Propagator Corrections}
 \label{fig:self}
\end{figure}

\begin{figure}[h!]
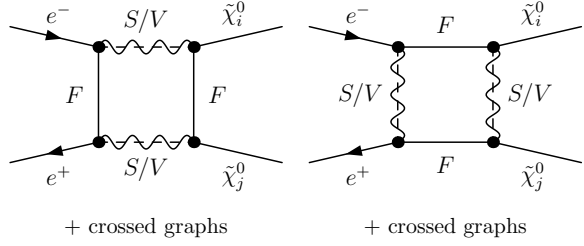

\hspace{2cm}
\mbox{\resizebox{200mm}{!}{
\begin{feynartspicture}(750,150)(4,1)
\FADiagram{+ crossed graphs}
\FAProp(0.,15.)(6.5,13.5)(0.,){/Straight}{1}
\FALabel(3.59853,15.2803)[b]{$e^-$}
\FAProp(0.,5.)(6.5,6.5)(0.,){/Straight}{-1}
\FALabel(3.59853,4.71969)[t]{$e^+$}
\FAProp(20.,15.)(13.5,13.5)(0.,){/Straight}{0}
\FALabel(16.4577,15.0367)[b]{$\tilde \chi_i^0$}
\FAProp(20.,5.)(13.5,6.5)(0.,){/Straight}{0}
\FALabel(16.4577,4.96329)[t]{$\tilde \chi_j^0$}
\FAProp(6.5,13.5)(6.5,6.5)(0.,){/Straight}{0}
\FALabel(5.43,10.)[r]{$F$}
\FAProp(6.5,13.5)(13.5,13.5)(0.,){/ScalarDash}{0}
\FAProp(6.5,13.5)(13.5,13.5)(0.,){/Sine}{0}
\FALabel(10.,14.57)[b]{$S/V$}
\FAProp(6.5,6.5)(13.5,6.5)(0.,){/ScalarDash}{0}
\FAProp(6.5,6.5)(13.5,6.5)(0.,){/Sine}{0}
\FALabel(10.,5.43)[t]{$S/V$}
\FAProp(13.5,13.5)(13.5,6.5)(0.,){/Straight}{0}
\FALabel(14.57,10.)[l]{$F$}
\FAVert(6.5,13.5){0}
\FAVert(6.5,6.5){0}
\FAVert(13.5,13.5){0}
\FAVert(13.5,6.5){0}

\FADiagram{+ crossed graphs}
\FAProp(0.,15.)(6.5,13.5)(0.,){/Straight}{1}
\FALabel(3.59853,15.2803)[b]{$e^-$}
\FAProp(0.,5.)(6.5,6.5)(0.,){/Straight}{-1}
\FALabel(3.59853,4.71969)[t]{$e^+$}
\FAProp(20.,15.)(13.5,13.5)(0.,){/Straight}{0}
\FALabel(16.4577,15.0367)[b]{$\tilde \chi_i^0$}
\FAProp(20.,5.)(13.5,6.5)(0.,){/Straight}{0}
\FALabel(16.4577,4.96329)[t]{$\tilde \chi_j^0$}
\FAProp(6.5,13.5)(6.5,6.5)(0.,){/ScalarDash}{0}
\FAProp(6.5,13.5)(6.5,6.5)(0.,){/Sine}{0}
\FALabel(5.43,10.)[r]{$S/V$}
\FAProp(6.5,13.5)(13.5,13.5)(0.,){/Straight}{0}
\FALabel(10.,14.57)[b]{$F$}
\FAProp(6.5,6.5)(13.5,6.5)(0.,){/Straight}{0}
\FALabel(10.,5.43)[t]{$F$}
\FAProp(13.5,13.5)(13.5,6.5)(0.,){/ScalarDash}{0}
\FAProp(13.5,13.5)(13.5,6.5)(0.,){/Sine}{0}
\FALabel(14.57,10.)[l]{$S/V$}
\FAVert(6.5,13.5){0}
\FAVert(6.5,6.5){0}
\FAVert(13.5,13.5){0}
\FAVert(13.5,6.5){0}

\end{feynartspicture}}}
\caption[box]
{Generic Box Corrections}
 \label{fig:box}
\end{figure}

The radiative corrections to the neutralino pair production include the following generic structure of one-loop Feynman diagrams: 
The virtual vertex corrections Fig.~\ref{fig:vertex}, the corrections to the $\tilde{e}_{L,R}$ and $Z^0$ propagators Fig.~\ref{fig:self}, 
and the box graph contributions Fig.~\ref{fig:box}. The notation $F$, $V$, and $S$ stand for all possible fermion, vector and scalar particles
in the MSSM, respectively. $U$ denotes the FP ghosts.  Diagrams with loops on the external fermion lines are included in the definition 
of the counter terms as wave function corrections.
In this work, the complete set of Feynman graphs is calculated with help of the packages FeynArts and FormCalc \cite{feyn}. We implemented
our renormalization procedure into these packages. 
For a proper treatment of the appearing UV divergencies, counter terms are introduced in the on-shell renormalization scheme.  
To preserve supersymmetry, the used regularization scheme is dimensional reduction ($\drbar$). 
The loop graphs with virtual photon exchange also introduce IR singularities. Therefore, real photon emission has to be included to obtain 
a finite result.
\begin{equation}
\sigma^{corr}(e^+e^- \rightarrow \cch_i \cch_j)= \sigma^{ren}(e^+e^- \rightarrow
\cch_i \cch_j)+\sigma(e^+e^- \rightarrow
\cch_i \cch_j\gamma)
\end{equation}
For the numerical analysis, we have also used the programs LoopTools and FF \cite{loopFF}.

\subsection{Renormalization}
\noindent{\bf Wave function counter terms}\\
In the prescription of the used on-shell renormalization scheme all involved fields get the following shifts to obtain the so-called
wave function corrections.
\begin{eqnarray}
 \hspace*{-10mm} \ch^0_i \!\! & \rightarrow (\d_{ij}+\onehf\d\tilde Z^L_{ij}P_L+\onehf\d\tilde Z^R_{ij}P_R)\ch^0_j\,,
& \hspace*{-2mm}
 \left(\!
\begin{array}{c}
 f_L \\ f_R
     \end{array}\!\right)\rightarrow \left(\begin{array}{cc}
1+\onehf\d Z_L & 0 \\ 0 & 1+\onehf\d Z_R
     \end{array}\right)\left(\!
\begin{array}{c}
 f_L \\ f_R
     \end{array}\!\right),\\
 \hspace*{-10mm} Z_\mu \!\! & \rightarrow (1+\onehf\d Z_{ZZ})Z_\mu+\onehf\d Z_{Z\g}A_\mu\,,
& \hspace*{-2mm}
\left(\!
\begin{array}{c}
 \sf_L \\ \sf_R
     \end{array}\!\right)\rightarrow \left(\begin{array}{cc}
1+\onehf\d Z^\sf_L & 0 \\ 0 & 1+\onehf\d Z^\sf_R
     \end{array}\right)
\left(\!
\begin{array}{c}
 \sf_L \\ \sf_R
     \end{array}\!\right)\,.
\end{eqnarray} 

with the definition of the renormalization constants

\begin{eqnarray}\label{waveconsts}
\d Z_{ZZ}~&=&~ - \Re\,\dot\Pi_{ZZ}(m_Z^2)\,,\;\quad \d Z_{Z\g}~=~
\frac{2\,\Re\,\Pi_{Z\g}(0)}{m_Z^2}\,,\\ 
\d Z^\sf_{L}~&=&~ - \Re\,\dot\Pi_{LL}^{\sf} (m_{\sf_{L}}^2)\,,\;\quad
\d Z^\sf_{R}~=~ - \Re\,\dot\Pi_{RR}^{\sf} (m_{\sf_{R}}^2)\,,\\ 
\d Z^L ~&=&~ \Re \bigg[
-\Pi^{L} (m_f^2)- m_f^2(\dot\Pi^{L} (m_f^2)+ \dot\Pi^{R}
(m_f^2))+\frac{1}{2 m_f}(\Pi^{SL} (m_f^2)-\Pi^{SR}
(m_f^2))\nonumber\\ && 
\hspace{2cm} - m_f(\dot\Pi^{SL}
(m_f^2)+\dot\Pi^{SR} (m_f^2))\bigg],\\
\d\tilde Z^L_{ii} ~&=&~ \Re \bigg[
-\Pi_{ii}^{L} (m_{\cch_i}^2)- m_{\cch_i}^2(\dot\Pi_{ii}^{L} (m_{\cch_i}^2)+ \dot\Pi_{ii}^{R}
(m_{\cch_i}^2))+\frac{1}{2 m_{\cch_i}}(\Pi_{ii}^{SL} (m_{\cch_i}^2)-\Pi_{ii}^{SR}
(m_{\cch_i}^2))\nonumber\\ && 
\hspace{2cm} - m_{\cch_i}(\dot\Pi_{ii}^{SL}
(m_{\cch_i}^2)+\dot\Pi_{ii}^{SR} (m_{\cch_i}^2))\bigg],\\
\d\tilde Z^L_{ij}~&=&~c_{ij}\, \Re\left[
m_{\cch_j}^2\Pi_{ij}^{L} (m_{\cch_j}^2)+m_{\cch_i}m_{\cch_j}\Pi_{ij}^{R}(m_{\cch_j}^2)
+m_{\cch_i}\Pi_{ij}^{SL}(m_{\cch_j}^2)+m_{\cch_j}\Pi_{ij}^{SR}(m_{\cch_j}^2)\right],
\non \hspace{-20mm}\\ 
\\
\d Z^R~&=&~\d Z^L(L \leftrightarrow R)\,,\;\quad
\d\tilde Z^R_{ii}~=~\d Z^L_{ii}(L \leftrightarrow R)\,,\;\quad
\d\tilde Z^R_{ij}~=~\d Z^L_{ij}(L \leftrightarrow R)\,,
\end{eqnarray}
where $\Pi_{(ij)}(k^2)=\ksla P_L\Pi_{(ij)}^L(k^2)+\ksla P_R\Pi_{(ij)}^R(k^2)+P_L\Pi_{(ij)}^{SL}(k^2)+P_R\Pi_{(ij)}^{SR}(k^2)$, 
$\dot\Pi (m^2)=\left[\frac{\partial}{\partial k^2}\Pi(k^2)\right]_{k^2=m^2 }$ and $c_{ij} = 2/(m_{\cch_i}^2-m_{\cch_j}^2)$.
For the neutralinos it holds $\Pi_{ij}^{R}(p^2)=\Pi_{ji}^{L}(p^2)\,,\Pi_{ij}^{SR/L}(p^2)=\Pi_{ji}^{SR/L}(p^2)$, because of their Majorana nature.
Since we neglect the selectron mixing, no sfermion mixing angle need to be renormalized.\\ \\

\noindent{\bf Neutralino and sfermion mass matrix renormalization}\\
In the MSSM, the four neutralino masses depend on the SUSY parameters $M'$, $M$, $\mu$, and $\tan\b$ and the SM parameters $m_Z$ and $\sin\theta_W$. 
As $M$, $\mu$ and $\tan\b$ also enter the chargino mass matrix, the renormalization of the neutralino, chargino and SM sectors is interrelated.
Therefore, it is necessary to take into account radiative corrections to the $\cch$ masses and the rotation matrix. 
For the on-shell renormalization two different approaches are essentially known in the literature, \cite{mcvienna} and \cite{mchollik, ghs}. 
Although the corrections in the neutralino masses are in general small, these shifts can lead to large effects near the threshold.  
It would be possible to adopt a renormalization scheme for each channel in such a way, 
that the two produced neutralinos are input parameters and do not obtain
mass corrections. A threshold shift would thus be avoided, but this leaves us with the problem that the renormalized processes have unequal counter terms
for different production channels, which would lead to different meanings of the neutralino and chargino mass parameters.
Here we use the on-shell scheme described in \cite{mcvienna}. We define an improved tree-level, 
where the process-independent mass matrix renormalization is already included and separated from the residual weak corrections.
Absorbing the finite correction $\D\ZN_{ij}$ to the rotation matrix $\ZN_{ij}$ in the improved tree-level is equivalent to defining an effective
coupling matrix $\ZN_{ij}+\D\ZN_{ij}$.\\
This yields the following counter terms for the neutralino mass matrix $\d Y_{ij}$ and the rotation matrix $\d \ZN_{ij}$.
\begin{eqnarray}
\d Y_{ij}&=&\frac{1}{2} \sum_{l,n=1}^4 \ZN_{li}\ZN_{nj}\,{\rm Re} \left[ m_{\cch_l}
\Pi^L_{nl} (m_{\cch_l}^2) + m_{\cch_n}
\Pi^R_{ln}(m_{\cch_n}^2)+\Pi^{SR}_{nl}(m_{\cch_l}^2)+\Pi^{SL}_{ln}(m_{\cch_n}^2)
\right]\,,\non\\ \\
\d \ZN_{ij} &=& \frac{1}{4}\sum_{k=1}^4\left(\d\tilde Z^L_{ik}- \d\tilde Z^R_{ki}\right)\ZN_{kj}
\,.
\end{eqnarray}
The same renormalization prescription can be applied to the sfermion sector. Counter terms for the SUSY breaking masses $M_{\tilde{Q},\tilde{L}}$ and
$M_{\tilde{U},\tilde{D},\tilde{E}}$, both entering the sfermion mass matrices, are introduced. 
Fixing $M_{\tilde{Q},\tilde{L}}$ in the down-type mass matrices results in a correction to the up-type masses and mixing angles \cite{MQprob}.
Hence, in our case, we have no additional corrections to the selectron masses. The 
correction to the electron sneutrino mass, which only appears in loop graphs, is of higher order and do not need to be considered.

\noindent{\bf Renormalization of the SM parameters}
\\ Since we use as input parameter for $\a$ the
$\overline{\rm MS}$ value at the $Z$ pole, $\a \equiv
\a(m_{\scriptscriptstyle Z})|_{\overline{\rm MS}} = e^2/(4\pi)$,
we get the counter term \cite{ChrisA0, 0111303}
\begin{eqnarray}\non
  \frac{\d e}{e} &=& \frac{1}{(4\pi)^2}\,\frac{e^2}{6} \Bigg[
  \,4 \sum_f N_C^f\, e_f^2 \bigg(\D + \log\frac{Q^2}{x_f^2} \bigg)
  + \sum_{\sf} \sum_{m=1}^2 N_C^f\, e_f^2
  \bigg( \D + \log\frac{Q^2}{m_{\sf_m}^2} \bigg)
  \\ \non
  && \hspace{18mm}
  +4 \sum_{k=1}^2 \bigg( \D + \log\frac{Q^2}{m_{\chp_k}^2} \bigg)
  + \sum_{k=1}^2 \bigg( \D + \log\frac{Q^2}{m_{H_k^+}^2} \bigg)
  - 22 \bigg( \D + \log\frac{Q^2}{m_{\scriptscriptstyle W}^2} \bigg)
  \Bigg] \,.
  \\
\end{eqnarray}
with $x_f = m_{\scriptscriptstyle Z} \ \forall\ m_f <
m_{\scriptscriptstyle Z}$ and $x_t = m_t$.  $N_C^f$ is the colour
factor, $N_C^f = 1, 3$ for (s)leptons and (s)quarks, respectively.
$\D$ denotes the UV divergence factor, \mbox{$\D = 2/\epsilon - \g
+ \log 4\pi$}.\\
The masses of the Z boson and the W boson are
fixed as the physical (pole) masses,
\begin{eqnarray}
\d m_Z^2 = \Re\,\Pi_{ZZ}(m_Z^2)\,,\qquad \d m_W^2 =
\Re\,\Pi_{WW}(m_W^2)\,,
\end{eqnarray}
and $\sin^2\theta_W$ is fixed by $\cos\theta_W=m_W/m_Z$.

\subsection{Definition of weak and QED corrections}
As mentioned before, the full one-loop corrections become IR convergent if also real photon emission is included in the calculation.
Because of these large additional corrections, it is desirable to treat the weak and QED parts separately. The easiest way 
to define pure ``weak corrections" would be to separate off all Feynman graphs with an additional photon attached to the tree-level diagrams. 
However, in our case this cannot be done in a gauge invariant and UV finite way due to the selectron exchange channels.
Another possibility would be to use the soft photon approximation \cite{Denner}, where only ``soft" photons up to a maximal energy $\D E$ are included: 
$\sigma^{weak}=\sigma^{soft}$ and $\sigma^{QED}=\sigma^{hard}$.
The weakness of this definition is the large $\D E$ dependence of the weak and QED components $\propto\log\frac{\D E^2}{s}$.
The sum of both is, however, cutoff independent.
Therefore, we extract the $\D E$ terms and the leading logarithms $\frac{\a}{\pi}L_e \equiv \frac{\a}{\pi}\log\frac{s}{m_e^2}$, caused by collinear
soft photon emission, from the weak corrections and add them to the QED corrections \cite{DennerQED}.
With this definition, both corrections are now $\D E$ independent.
The main part of the QED corrections arises from these leading logarithms $L_e$, originating from photons in beam direction.
This leads to a large dependence on experimental cuts and
detector specifications. We therefore use the structure
function formalism \cite{SFF} and subtract the leading logarithmic $O(\a)$ terms of the initial state radiation, $\sigma^{ISR,LL}(s)$. 
After subtraction of these process-independent terms, only the non-universal QED corrections remain. 
This gives for the total cross section the final expression:
\begin{eqnarray}
\sigma^{total}(s)~&=&~\sigma^{tree}(s)+\sigma^{weak}(s)+\sigma^{QED}(s)\,,\\
\sigma^{weak}(s)~&=&~\sigma^{soft}(s)+\frac{\a}{\pi} \left( (1-L_e)\log\frac{\D E^2}{s}-\frac{3}{2}L_e \right) \sigma^{tree}(s)\,,\\
\sigma^{QED}(s)~&=&~\sigma^{hard}(s)-\frac{\a}{\pi}\left( (1-L_e)\log\frac{\D E^2}{s}-\frac{3}{2}L_e \right)
\sigma^{tree}(s)-\sigma^{ISR,LL}(s)\,,
\end{eqnarray}
with
\begin{eqnarray}
\sigma^{ISR,LL}(s)~&=&~\frac{\a}{\pi}L_e\int_0^1dx\,\,\Phi(x)\,\sigma^{tree}(xs)\,,\\
\Phi(x)~&=&~\lim_{\e\rightarrow 0}\{\d(1-x)[\frac{3}{2}+2\log(\e)] +\theta(1-x-\e)\frac{1+x^2}{1-x}\}\,.
\end{eqnarray}
Further improvements would be to consider a more realistic electron spectrum and incorporate beamstrahlung in the calculations. 
Due to their strong dependence on the actual experimental conditions, we do not include these effects.

\section{Numerical results}\label{numericalresults}
For the numerical analysis, we concentrate on the production channels
\bea
e^+e^- \rightarrow \cch_1\cch_2  \qquad \rm{and} \qquad  e^+e^- \rightarrow \cch_2\cch_2. \nonumber
\eea
They are of special interest for future experiments because of their decay products and for kinematical reasons \cite{tesla}.
Due to the tree-level coupling structure, we study here two different scenarios: In the higgsino scenario the two lightest neutralinos
are both nearly pure higgsinos and therefore the process is dominated by the s-channel $Z^0$ exchange. In the gaugino scenario 
with a bino and a wino as $\cch_1$ and $\cch_2$ states, the selectron exchange diagrams play the most important role. In the following, we distinguish
between the naive tree-level, the improved tree-level with the corrections to the neutralino masses $m_{\cch_i}$ and the rotation matrix $\ZN_{ij}$
included, and the conventional weak and QED corrections to the improved tree-level as discussed in the last chapter. 
For the SM input parameters we use $\a(m_Z)=1/127.922$, $m_Z=91.1876$ GeV, and $m_W=80.423$ GeV.
\subsection{Higgsino scenario}
\begin{figure}[h!]
\begin{center}
\mbox{\resizebox{84mm}{!}{\includegraphics{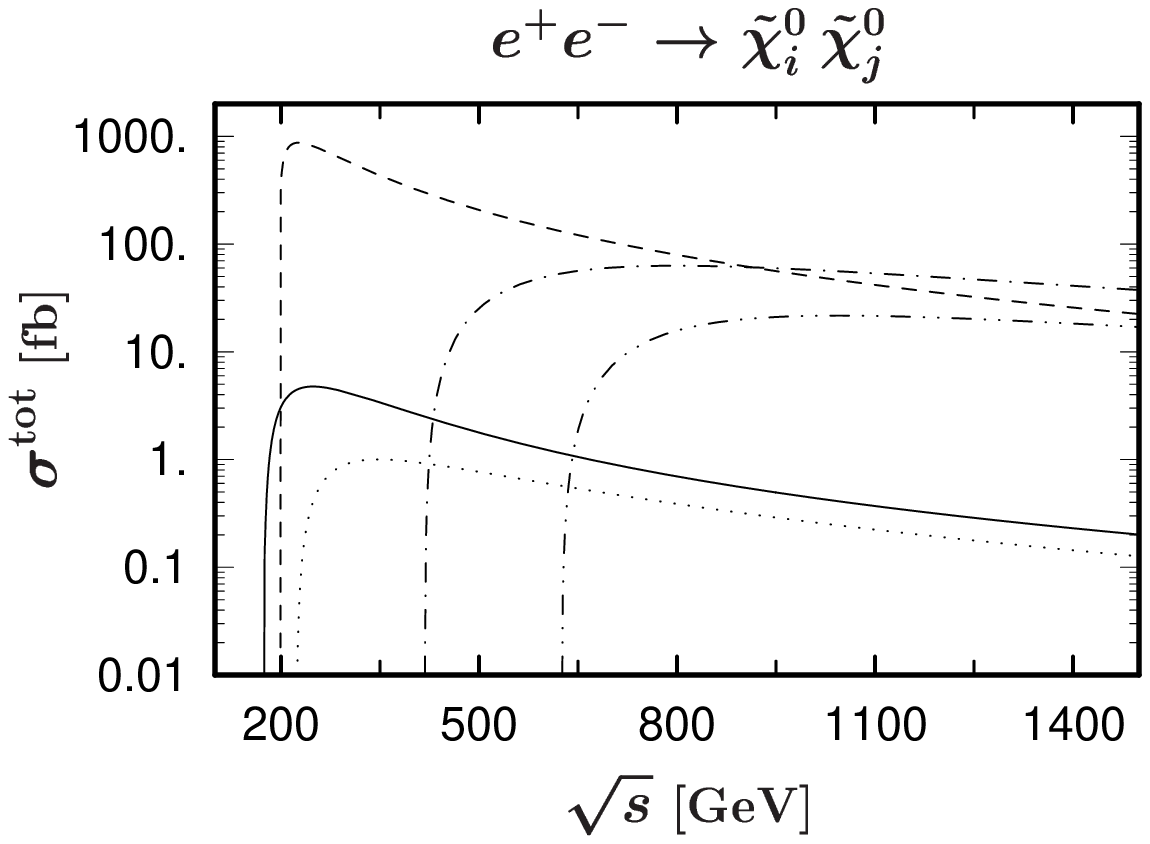}}}
\end{center}
\caption{Neutralino pair production in the naive tree-level approximation with \{full, dashed, dotted, dash-dotted, dash-dot-dotted\}=
\{$\cch_1\cch_1$, $\cch_1\cch_2$, $\cch_2\cch_2$, $\cch_3\cch_3$, $\cch_3\cch_4$\}.}
\label{fig:hlog}
\end{figure}

\begin{figure}[h!]
 \begin{center}
 \hspace*{-1mm}
 \mbox{\mbox{\resizebox{84mm}{!}{\includegraphics{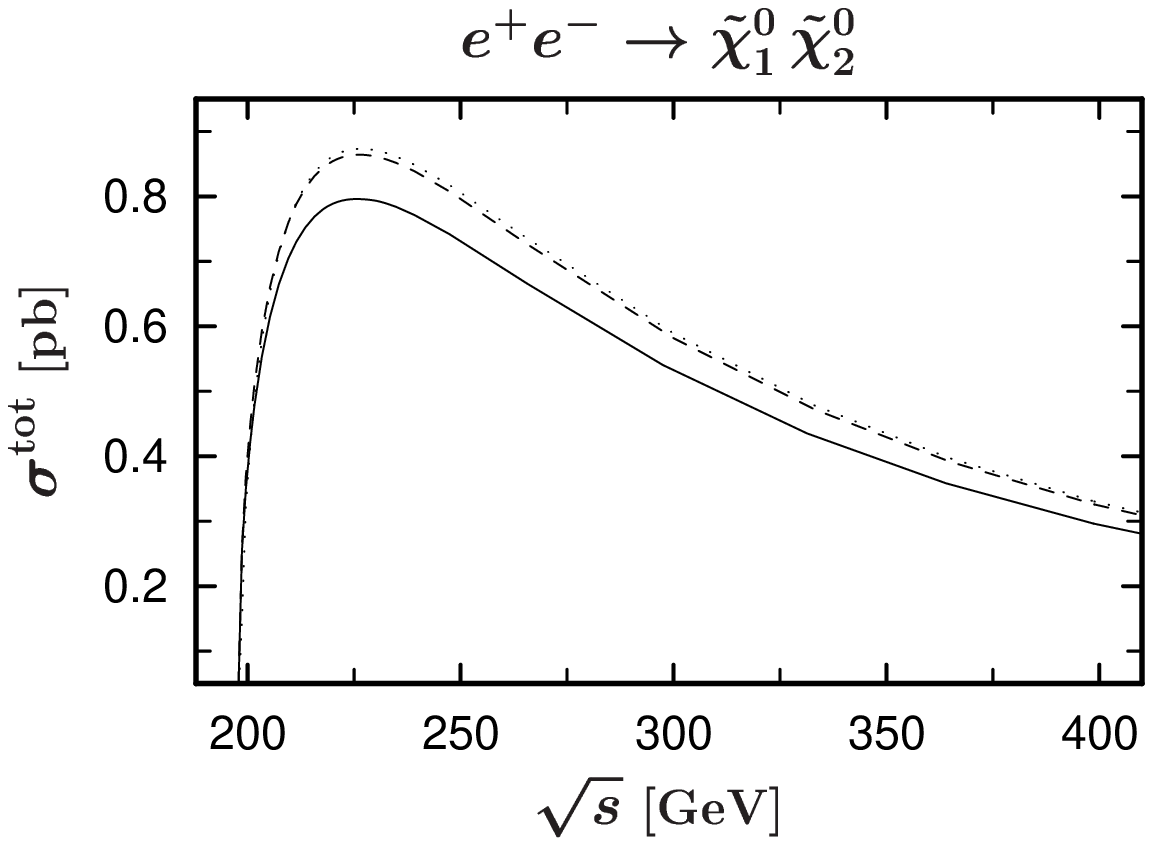}}} \hspace{-5mm}
 \mbox{\resizebox{84mm}{!}{\includegraphics{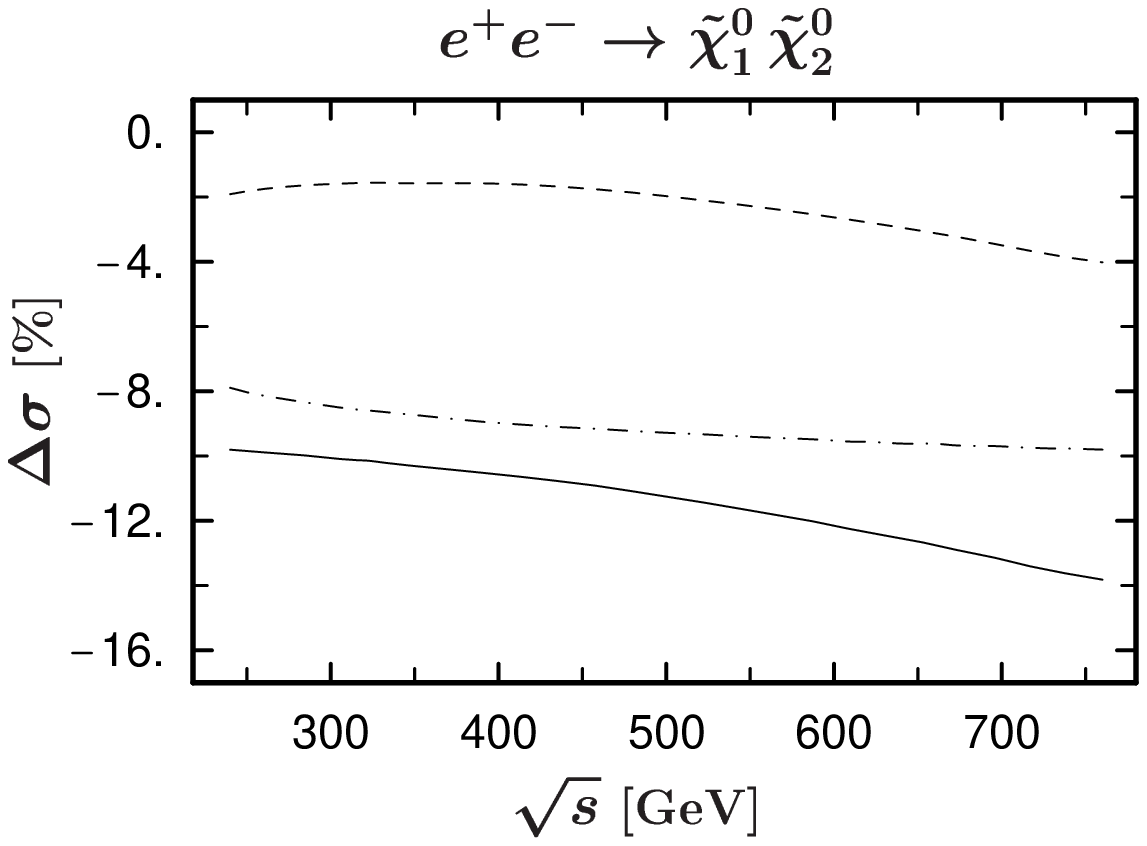}}}}
 \caption[fig6]
 {Corrections to the $\cch_1\cch_2$ higgsino scenario. Left: The total cross-section in the naive tree-level approximation
(dotted line), the improved tree-level (dashed line), and the full $O(\a)$ corrected without ISR (solid line). 
Right: The full $O(\a)$ without ISR (solid line), weak (dashed line) and
non-universal QED (dash-dotted line) corrections relative to the improved tree-level.}
 \label{fig:hcorr12}
 \end{center}
 \vspace{-7mm}
 \end{figure}

\begin{figure}[h!]
 \begin{center}
 \hspace*{-1mm}
 \mbox{\mbox{\resizebox{84mm}{!}{\includegraphics{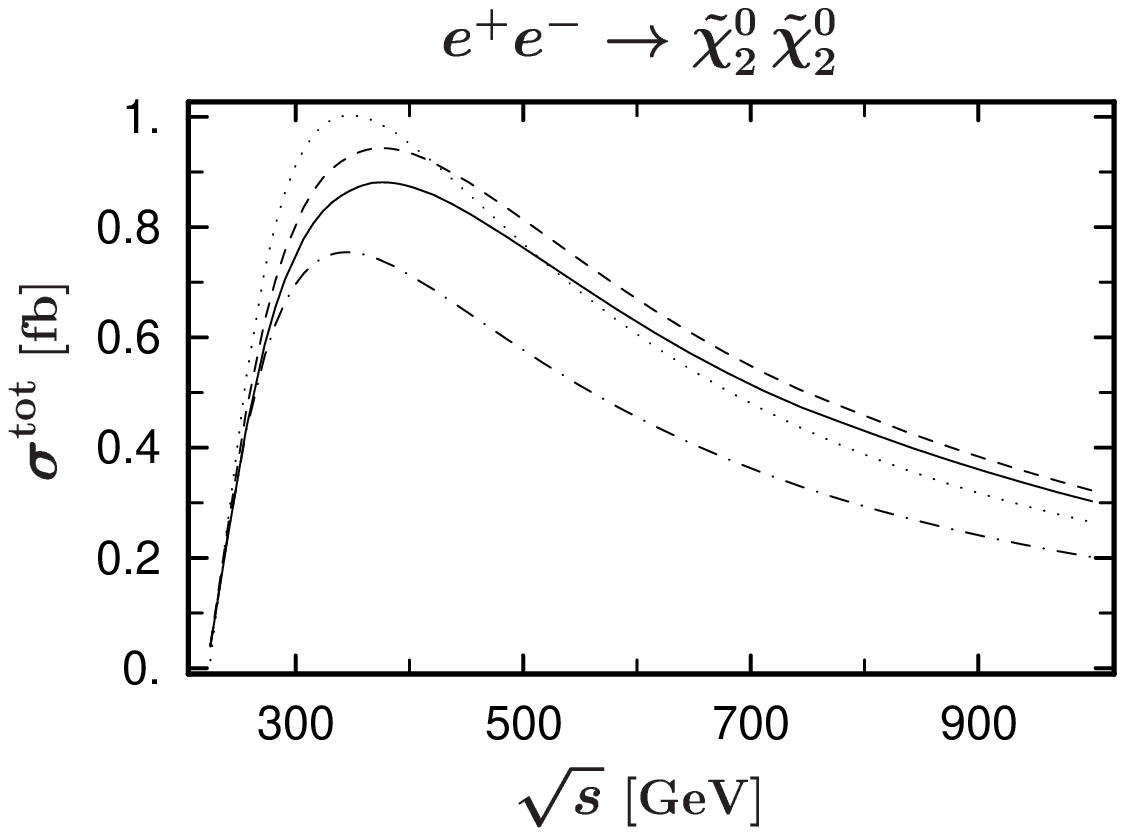}}} \hspace*{-5mm}
 \mbox{\resizebox{84mm}{!}{\includegraphics{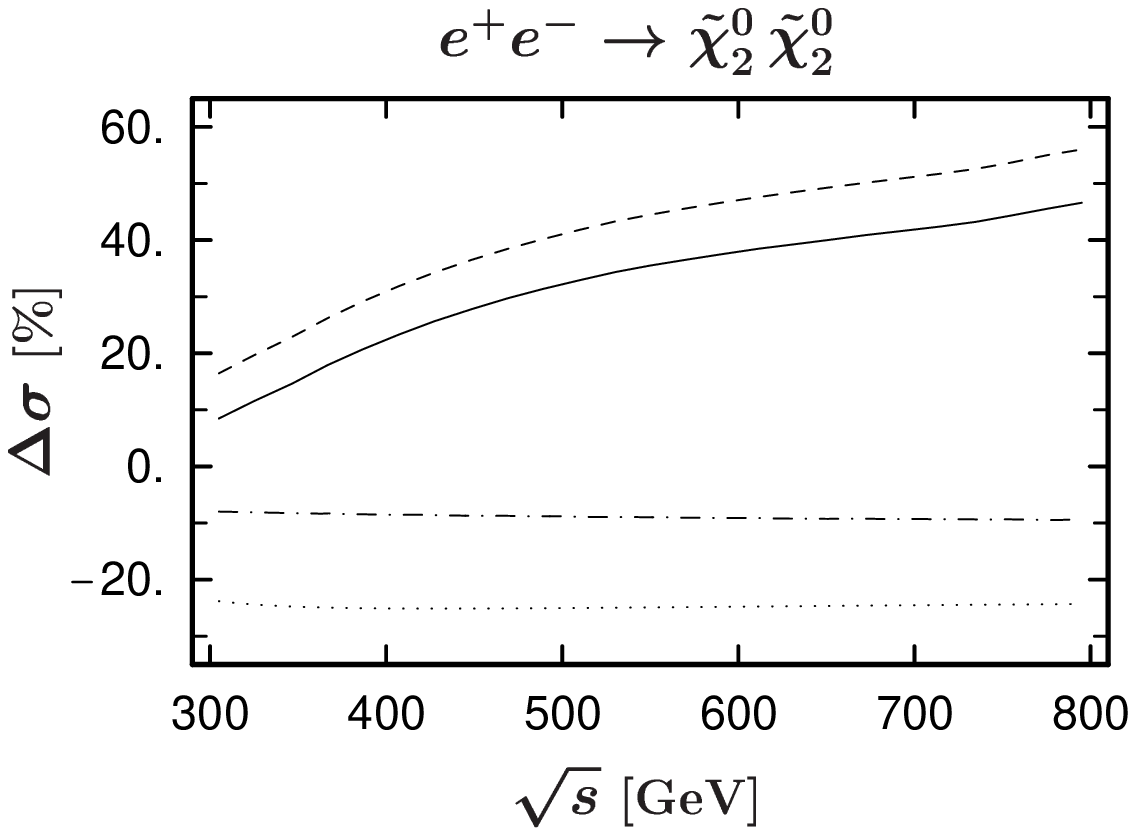}}}}
 \caption[fig6]
 {Corrections to the $\cch_2\cch_2$ higgsino scenario. Left: The total cross section in the naive tree-level approximation
(dotted line), the improved tree-level (dash-dotted line), with the weak corrections (dashed line), and the full $O(\a)$ corrected 
one without ISR (solid line). Right: The full $O(\a)$ without ISR (solid line), weak (dashed line) and
non-universal QED (dash-dotted line) corrections relative to the improved tree-level. The dotted line shows the effect of the 
mass matrix corrections relative to the naive tree-level.}
 \label{fig:hcorr22}
 \end{center}
 \vspace{-7mm}
 \end{figure}

For the definition of the higgsino scenario we use the following MSSM on-shell parameters in the convention \cite{mcvienna}:\\
$\tan\beta$~=~10;  $\mu$~=~-100 GeV; $M_2$~=~$2M_1$~=~400 GeV; $M_{Q,L}$~=~$M_{U,D,E}$~=~350 GeV; $A_f$~=~400 GeV; $M_{A^0}$~=~700 GeV.\,
This gives the one-loop corrected neutralino masses:\\
$\cch_1$ (94\% higgsino): 87.8 GeV \quad $\cch_2$ (97\% higgsino): 110.0 GeV\\
$\cch_3$ (94\% bino): 209.4 GeV \qquad $\cch_4$ (96\% wino): 415.2 GeV\\
In Fig.~\ref{fig:hlog}, we show the naive tree-level cross section for five different channels. The double higgsino production  
$e^+e^- \rightarrow \cch_i\cch_i$ with $i=1,2$ is highly suppressed due to the behaviour of the $Z^0\cch_i\cch_j$ coupling.\\
Numerical results for the radiative corrections to the $\cch_1\cch_2$ production are given in Fig.~\ref{fig:hcorr12}. The total non-universal
weak and QED corrections are in the range of -12\% in the investigated parameter region and thus have to be taken into account in future experiments.
In the case of $\cch_2\cch_2$ production, Fig.~\ref{fig:hcorr22}, the small tree-level $Z^0\cch_2\cch_2$ coupling leads
to an enhancement of the corresponding vertex corrections and to large box graph contributions. 
For the same reason, owing to the neutralino rotation matrix 
correction $\ZN_{ij}$ the effect is also highly increased. Therefore, there is a big difference between the naive and improved tree-level cross section.

\newpage
\subsection{Gaugino scenario}
\begin{figure}[h!]
 \begin{center}
 \mbox{\resizebox{84mm}{!}{\includegraphics{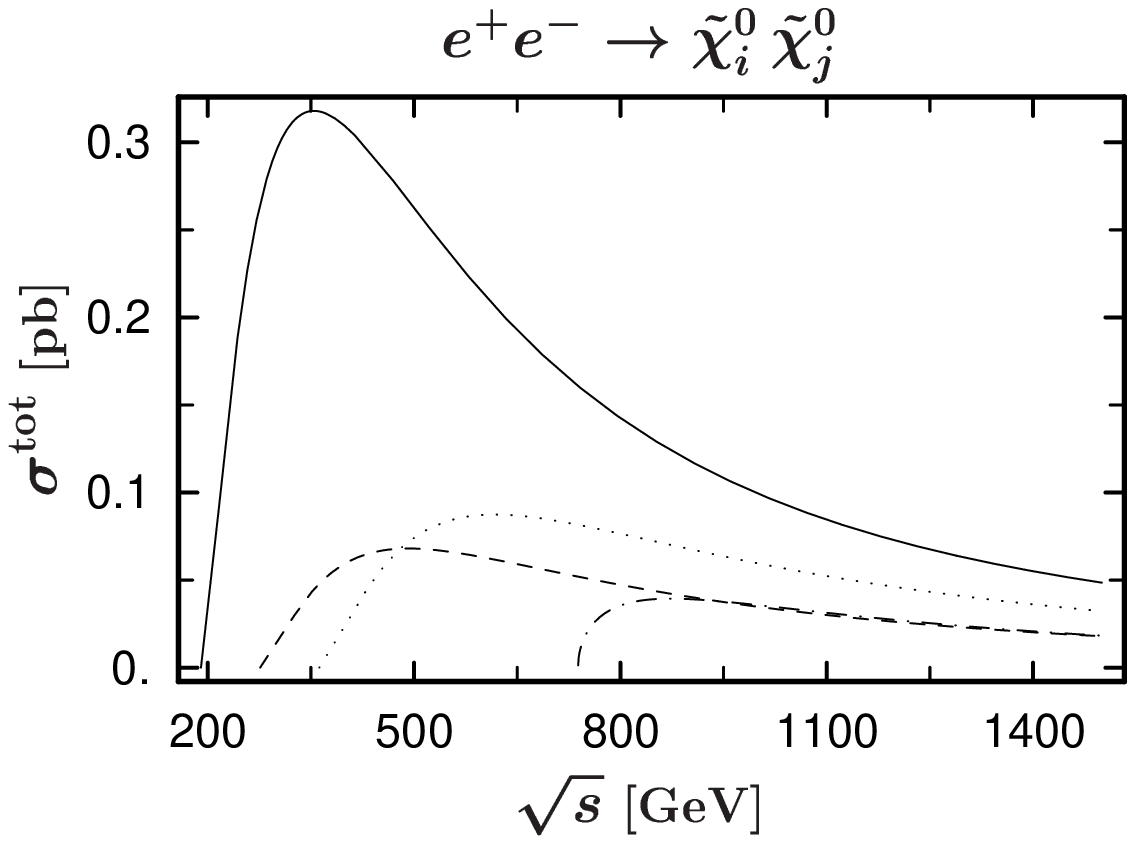}}}
 \end{center}
 \caption{Neutralino pair production in the naive tree-level approximation with 
 \{full, dashed, dotted, dash-dotted\}=\{$\cch_1\cch_1$, $\cch_1\cch_2$, $\cch_2\cch_2$, $\cch_3\cch_4$\}.}
\label{fig:gtree}
\end{figure}

\begin{figure}[h!]
 \begin{center}
 \hspace*{-1mm}
 \mbox{\mbox{\resizebox{84mm}{!}{\includegraphics{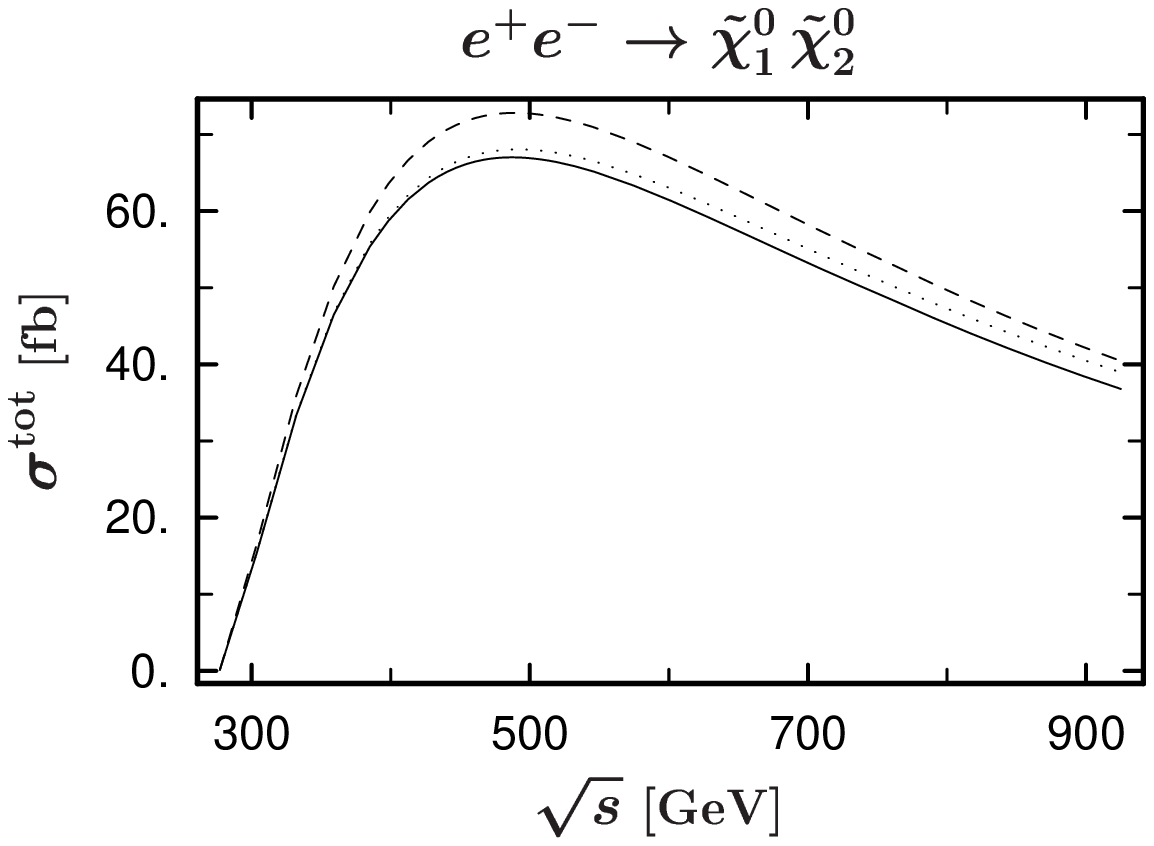}}} \hspace*{-5mm}
 \mbox{\resizebox{84mm}{!}{\includegraphics{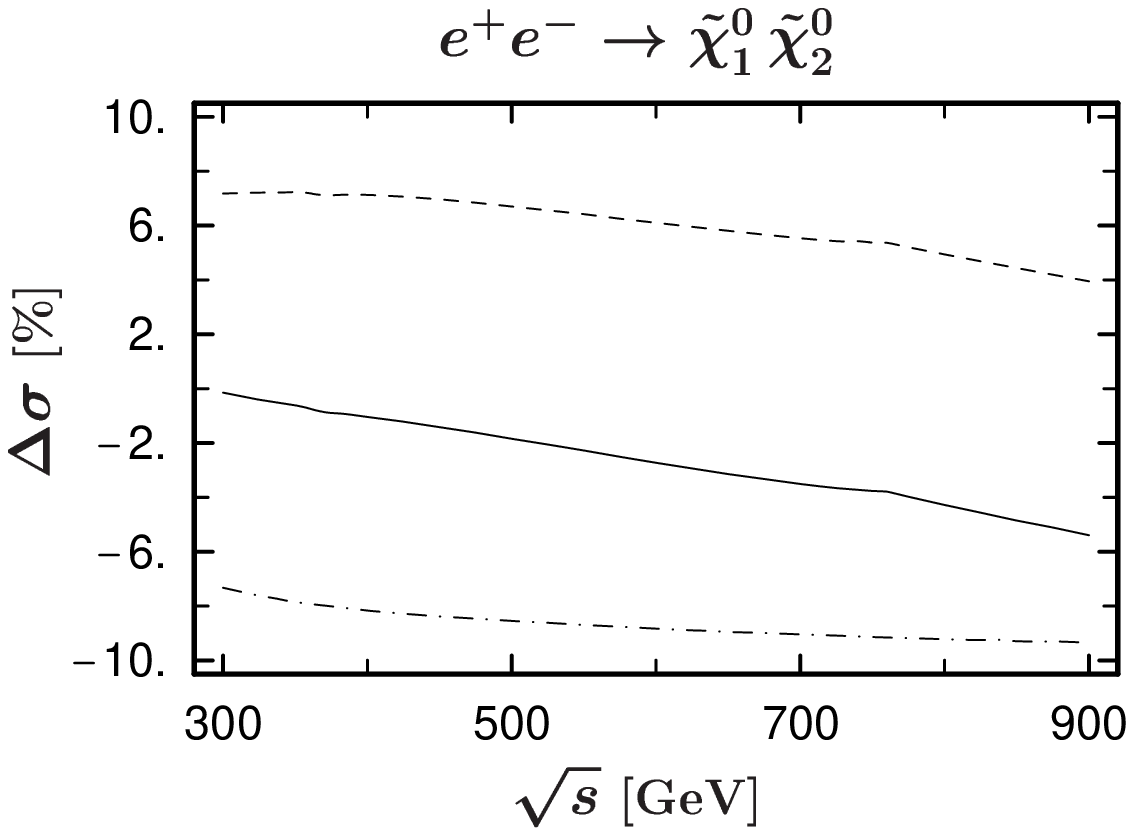}}}}
 \caption[fig6]
 {Corrections to the $\cch_1\cch_2$ SPS1a scenario. Left: The total cross-section in the naive tree-level approximation
(dotted line) and with weak (dashed line), and full $O(\a)$ (solid line) corrections without ISR. 
Right: The full $O(\a)$ without ISR (solid line), weak (dashed line) and
non-universal QED (dash-dotted line) corrections relative to the improved tree-level.}
 \label{fig:gcorr12}
 \end{center}
 \vspace{-7mm}
 \end{figure}

\begin{figure}[h!]
 \begin{center}
 \hspace*{-1mm}
 \mbox{\mbox{\resizebox{84mm}{!}{\includegraphics{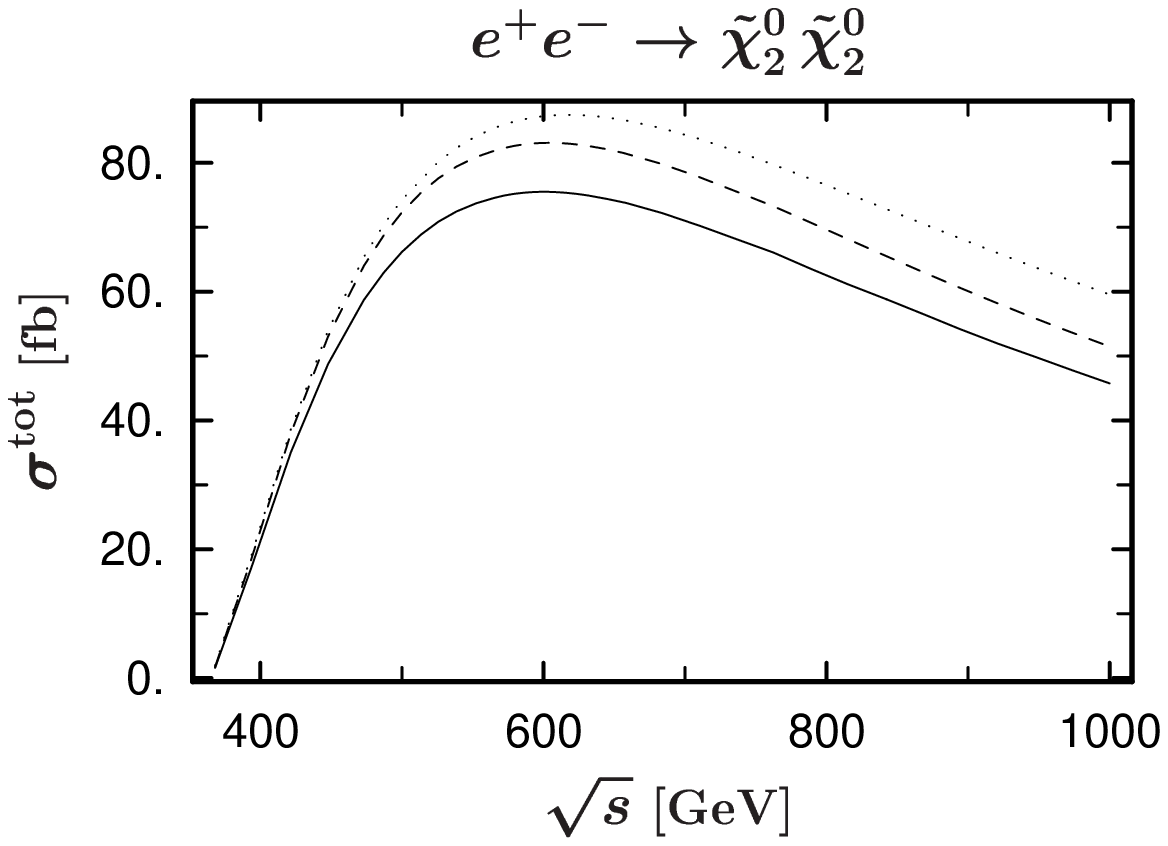}}} \hspace*{-5mm}
 \mbox{\resizebox{84mm}{!}{\includegraphics{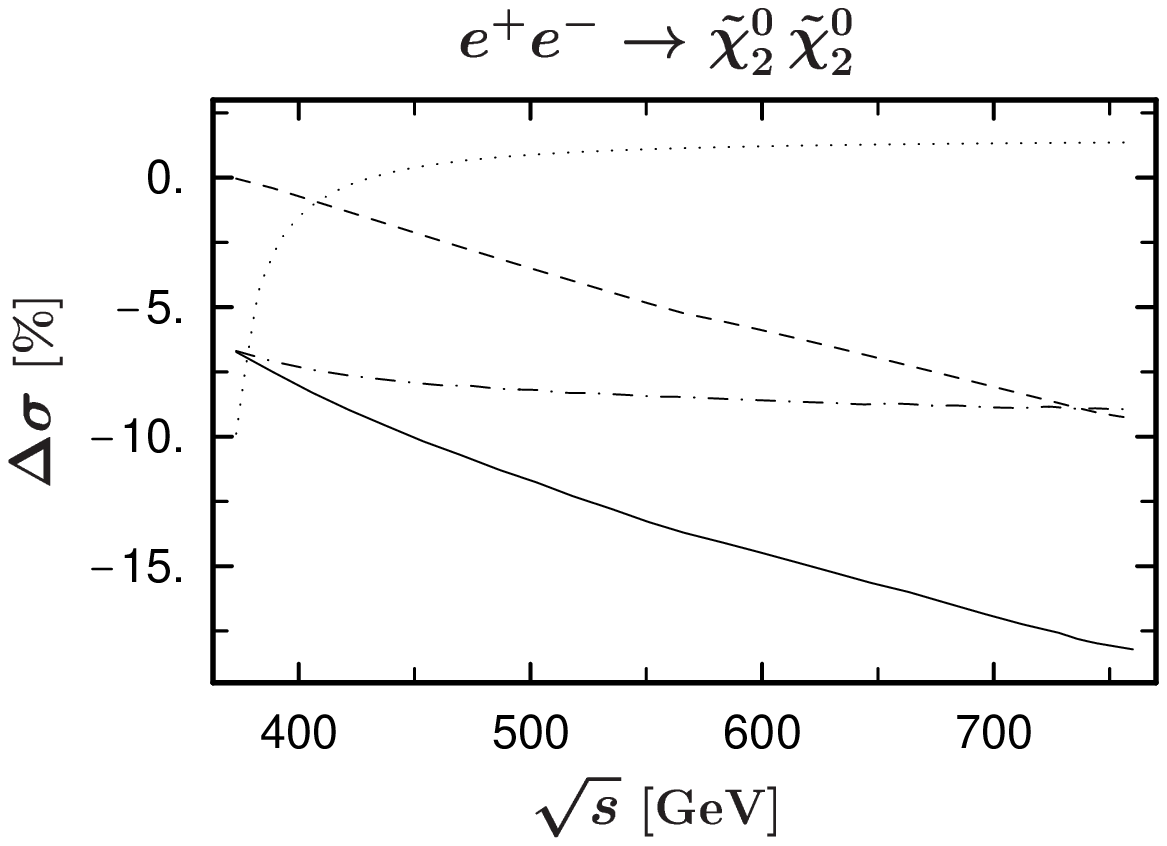}}}}
 \caption[fig6]
 {Corrections to the $\cch_2\cch_2$ SPS1a scenario. Left: The total cross-section in the naive tree-level approximation
(dotted line) and with weak (dashed line), and full $O(\a)$ (solid line) corrections without ISR.
Right: The full $O(\a)$ without ISR (solid line), weak (dashed line) and
non-universal QED (dash-dotted line) corrections relative to the improved tree-level. The dotted line shows the effect of the 
mass matrix corrections relative to the naive tree-level.}
 \label{fig:gcorr22}
 \end{center}
 \vspace{-7mm}
 \end{figure}

In the case of the gaugino scenario, we use as input the SPS1a \drbar benchmark values \cite{spsvalues}, 
defined at the scale Q = 454.7 GeV. With these values, we can calculate our on-shell parameters in a consistent way by subtraction
of the corresponding counter terms, e.g. $M_1=M_1^{\drbar}(Q)-\d Y_{11}(Q)$, and obtain:\\
$\tan\beta$~=~10.2;  $\mu$~=~353.1 GeV; $M_1$~=~97.9 GeV; $M_2$~=~197.6 GeV; $M_{A^0}$~=~393.6 GeV.\\
In the sfermion sector, we only need the selectron mass parameters: $M_{L}$~=~198.0 GeV; $M_{E}$~=~138.0 GeV. For all other parameters,
we can use the \drbar or on-shell values. The differences are of higher order for our calculation.  
For the neutralino states we get:\\
$\cch_1$ (97\% bino): 94.8 GeV \qquad\quad $\cch_2$ (88\% wino): 181.5 GeV\\
$\cch_3$ (99\% higgsino): 360.3 GeV \quad $\cch_4$ (88\% higgsino): 377.4 GeV\\
Note that the SPS1a scenario is defined by \drbar parameter values. Thus the one-loop on-shell parameters given here can differ from those calculated
in other renormalization schemes. The on-shell masses are of course the same up to higher orders.\\ 
In Fig.~\ref{fig:gtree}, we show the tree-level cross section for all three possible gaugino production channels and the higgsino
$\cch_3\cch_4$ production. The double higgsino channel $e^+e^- \rightarrow 2\cch_i$ with i=3,4 or mixed gaugino-higgsino channels are suppressed
due to the given coupling structure.
The full $O(\a)$ radiative corrections for the $\cch_1\cch_2$ production, given in Fig~\ref{fig:gcorr12}, are only in the few percent range because
of the cancellation between the weak and QED corrections, especially near the threshold.
While the QED corrections for the $\cch_2\cch_2$ channel, see Fig.~\ref{fig:gcorr22}, are also moderate and show a similar behavior to the previous case, 
the weak corrections strongly depend on $\sqrt{s}$. 
For large $\sqrt{s}$ this can be studied in the so-called Sudakov approximation \cite{beccaria}.
The corrections are -10\% at $\approx 750$ GeV and even larger at higher energies.
One reason is that the $\cch_2$, being mainly a wino, has also an 11\% higgsino component, which effects the weak corrections in a similar way to the 
$\cch_2$ pair production in the higgsino scenario. This results in a large negative correction for the sum of the QED and weak part. 

\section{Conclusions}\label{conclusions}
We have calculated the full one-loop elektroweak corrections to the neutralino pair production in $e^+e^-$ collisions.
The chosen renormalization scheme can be used for the complete MSSM parameter space and all production channels 
$e^+e^- \rightarrow \cch_i\cch_j$ with i,j = 1,2,3,4. The process independent corrections to the neutralino mass matrix
are included in the definition of an improved tree level. We paid particular attention to an appropriate definition of weak and QED corrections.
We extracted the non-universal QED corrections by subtracting the initial state radiation (ISR).
The full one-loop corrections without ISR are in the range of 5-20\% and in some cases even larger and thus have to be taken into account
in future linear collider experiments. 
\section*{Acknowledgements}
The authors acknowledge support from EU under the
HPRN-CT-2000-00149 network programme. The work was also supported
by the ``Fonds zur F\"orderung der wissenschaftlichen Forschung'' of Austria,
project no. P13139-PHY.


\begin{thebibliography}{99}

\bibitem{MSSM}
H.~P.~Nilles, Phys. Rep. {\bf 110} (1984) 1; H.~E.~Haber and
G.~L.~Kane, Phys. Rep. {\bf 117} (1985) 75; R.~Barbieri, Riv. Nuov. Cim. {\bf 11} (1988) 1.

\bibitem{tesla}
TESLA Technical Design Report, Part III, Eds.: R. D. Heuer, D.
Miller, F. Richard, and P. M. Zerwas, DESY 2001-011.

\bibitem{lincol}
C.~Adolphsen, et al., International Study Group Collaboration,
International study group progress report on linear collider
development, SLAC-R-559 and KEK-REPORT-2000-7.

\bibitem{BlankHollik}
T.~Blank, W.~Hollik [arXiv:hep-ph/0011092].

\bibitem{Freitas}
A.~Freitas, D.~J.~Miller, P.~M.~Zerwas, Eur. Phys. J. {\bf C21}
(2001) 361-368 [arXiv:hep-ph/0106198]; A.~Freitas, A.~von
Manteuffel, P.~M.~Zerwas, [arXiv:hep-ph/0310182].

\bibitem{ArhribHollik}
A.~Arhrib, W.~Hollik [arXiv:hep-ph/0311149], to appear in
Eur. Phys. J. C.

\bibitem{KarolChris}
K.~Kovarik, C.~Weber, H.~Eberl, W.~Majerotto [arXiv:hep-ph/0401092],
to appear in Phys. Lett. B.

\bibitem{ghs}
J.~Guasch, W.~Hollik, J.~Sola, JHEP 0210:040 (2002) [arXiv:hep-ph/0207364].
 
\bibitem{ChrisA0}
C.~Weber, H.~Eberl, W.~Majerotto Phys. Rev. {\bf D68} (2003) 093011 [arXiv:hep-ph/0308146];
C.~Weber, H.~Eberl, W.~Majerotto Phys. Lett. {\bf B572} (2003) 56  [arXiv:hep-ph/0305250]. 

\bibitem{mcvienna}
H.~Eberl, M.~Kincel, W.~Majerotto, Y.~Yamada, Phys. Rev. {\bf D64} (2001) 115013 [arXiv:hep-ph/0104109];\\
W.~\"Oller, H.~Eberl, W.~Majerotto, C.~Weber, Eur. Phys. J. {\bf C29} (2003) 563 [arXiv:hep-ph/0304006].

\bibitem{mchollik}
T.~Fritzsche, W.~Hollik, Eur. Phys. J. {\bf C24} (2002) 619 [arXiv:hep-ph/0203159].

\bibitem{feyn}
J.~K\"ublbeck, M.~B\"ohm, A.~Denner, Comput. Phys. Commun. {\bf
60} (1990) 165;\\ T.~Hahn, Comput. Phys. Commun. {\bf 140} (2001)
418;\\ T.~Hahn, C.~Schappacher, Comput. Phys. Commun. {\bf 143}
(2002) 54;\\ T.~Hahn, M.~Perez-Victoria, Comput. Phys. Commun.
{\bf 118} (1999) 153.

\bibitem{loopFF}
G.~J.~Oldenborgh, Comput. Phys. Commun. {\bf 66} (1991) 1;\\
T.~Hahn, Acta Phys. Polon. {\bf B30} (1999) 3469.

\bibitem{MQprob}
A.~Bartl et al., Phys. Lett. {\bf B402} (1997) 303.

\bibitem{0111303}
H.~Eberl, M.~Kincel, W.~Majerotto and Y.~Yamada, Nucl. Phys. {\bf
B625} (2002) 372 [arXiv:hep-ph/0111303].

\bibitem{Denner}
A.~Denner, Fortschr. Phys. {\bf 41} (1993) 307.


 
\bibitem{DennerQED}
A.~Denner, S.~Dittmaier, Nucl. Phys {\bf B398} (1993) 239;\\
M.~B\"ohm, S.~Dittmaier, Nucl. Phys {\bf B409} (1993) 3.

\bibitem{SFF}
W.~Beenakker, A.~Denner Int. J. Mod. Phys. {\bf A9} (1994) 4837 and references therein.

\bibitem{spsvalues}
B.~C.~Allanach et al., Eur. Phys. J. {\bf C25} (2002) 13 [arXiv:hep-ph/0202233].

\bibitem{beccaria}
M.~Beccaria, F.~M.~Renard, C.~Verzegnassi, [arXiv:hep-ph/0203254];
M.~Beccaria, M.~Melles, F.~M.~Renard, S.~Trimarchi, C.~Verzegnassi, Int. J. Mod. Phys.
{\bf A18} (2003) 5069 [arXiv:hep-ph/0304110].

\end{thebibliography}
\end{document}